\documentclass[11pt]{article}
\usepackage{epsf}

\catcode`\@=11

\ifcase \@ptsize
    
\or 
\or 
\fi
\advance\textheight by \topskip

\catcode`\@=12

\newcommand{\beqn}{\begin{eqnarray}}
\newcommand{\eeqn}{\end{eqnarray}}
\newcommand{\beqns}{\begin{eqnarray*}}
\newcommand{\eeqns}{\end{eqnarray*}}

\newcommand{\rmd}{\mbox{d}}
\newcommand{\rme}{\mbox{e}}
\newcommand{\rmi}{\mbox{i}}

\newcommand{\dd}[2]{{\rmd{#1}\over\rmd{#2}}}

\newcommand{\wt}{\widetilde}

\newcommand{\bfit}[1]{\mbox{\protect\boldmath $#1$}}

\catcode`\@=11
\newcommand{\ccite}[1]
{\@ifundefined{b@#1}{\bf ?}{\@nameuse{b@#1}}}
\catcode`\@=12

\newcounter{remark}

\begin{document}

\begin{center}
{\Large\bf The fourth virial coefficient of anyons}\\[0.5cm]

{\large
Anders Kristoffersen$^{\mbox{\scriptsize\ a}}$,
Stefan Mashkevich$^{\mbox{\scriptsize\ a,b,c}}$,
Jan Myrheim$^{\mbox{\scriptsize\ a,c}}$,
K{\aa}re Olaussen$^{\mbox{\scriptsize\ a,c}}$\\}
\ \\
{\large\it
$^{\mbox{\scriptsize a}}$
Department of Physics,
The Norwegian University of Science and Technology,\\
N--7034 Trondheim, Norway\\
$^{\mbox{\scriptsize b}}$
Bogolyubov Institute for Theoretical Physics,\\
252143 Kiev, Ukraine\\
$^{\mbox{\scriptsize c}}$
Centre for Advanced Study, Norwegian Academy of Science and Letters,\\
Drammensveien 78, N--0271 Oslo, Norway
}

\end{center}

\begin{abstract}
We have computed by a Monte Carlo method the fourth virial
coefficient of free anyons,
as a function of the statistics angle $\theta$.
It can be fitted by a four term Fourier series, in which
two coefficients are fixed by the known
perturbative results at the boson and fermion points.
We compute partition functions by means of path integrals,
which we represent diagrammatically in such a way that
the connected diagrams give the cluster coefficients.
This provides a general proof that all cluster and virial
coefficients are finite.
We give explicit polynomial approximations for all path integral
contributions to all cluster coefficients, implying that only
the second virial coefficient is statistics dependent,
as is the case for two-dimensional exclusion statistics.
The assumption leading to these approximations
is that the tree diagrams dominate and factorize.

\end{abstract}

\section{Introduction}

Anyons are identical particles in the plane characterized
by a statistics phase angle $\theta$
\nocite{JML:JM:idpart1,GMS:idpart1,FW:idpart1}
{[\ccite{JML:JM:idpart1}--\ccite{FW:idpart1}]}.
They may be thought of
e.g.\ as bosons with a ``statistics interaction'' between
two particles which is an angular vector potential proportional to
$\theta/r$ at a relative distance $r$. Since this is
apparently a long range interaction, it is not immediately
obvious that the cluster and virial coefficients should be finite
in the thermodynamic limit.

We work out here a path integral representation for the cluster
coefficients of anyons, which shows that they are finite.
{}From this point of view the statistics interaction has short
range, thus the important property is not the $1/r$ dependence
of the vector potential but rather the pointlike
nature of the flux. Its range is temperature dependent, however,
because it is effective when the particle
paths wind around each other, and each path in the path integral
represents Brownian motion of a particle in the plane, covering
an area inversely proportional to the temperature.

The path integral representation for the cluster coefficients is
in fact quite general, and can be applied to anyons in two dimensions,
as well as to bosons and fermions in any dimension, interacting by
general scalar and vector potentials. The same argument for finiteness
holds in general, when the interaction range is sufficiently short.

We have computed numerically the fourth virial coefficient of free
anyons, by Monte Carlo evaluation of the four-particle path integrals.
We find that it is very nearly constant, i.e.\ nearly
equal to zero for all values of $\theta$. Thus the anyon
system is an approximate realization of Haldane's so-called
exclusion statistics \cite{FDMH91}, characterized by a continuously
variable parameter $g$, for which only the second virial
coefficient depends on $g$, in two dimensions \cite{SBI:DPA:JM:APP}.
The correspondence between the two kinds of fractional
statistics is given by the relation
\beqn
g=1-(1-\alpha)^2\;,
\label{g-alpha}
\eeqn
where $\alpha$ is the periodic function of $\theta$ defined in
eq.~(\ref{alpha_defined}) below.
This correspondence is only approximate, in fact it is known from
perturbation theory that the higher virial coefficients of anyons
all have a second order variation with $\theta$ at the boson and
fermion points
\nocite{AC:YG:SO,JMcC:SO:1,AC:JMcC:SO:2,SenIII,AdV:SO:esso}
\nocite{AdV:SO:2,MS:JJMV:IZ:a3}
{[\ccite{AC:YG:SO}--\ccite{MS:JJMV:IZ:a3}]}.
Our results are consistent with perturbation theory, within the
precision obtained.

All observable properties of anyons must be periodic functions
of $\theta$ with period $2\pi$. Energy eigenvalues and
eigenfunctions are analytic functions of $\theta$,
except that some are non-analytic
at $\theta=0$, varying like $|\theta|$ rather than $\theta$.
Hence the partition functions and all thermodynamic quantities
derivable from them will be analytic functions of $\theta$,
even at the fermion point $\theta=\pi$, but generally not at
the boson point $\theta=0$.

In the absence of an external magnetic field, the theory is both
time reversal and parity invariant if each of these transformations
is defined so as to include a change in sign of $\theta$.
It follows that energy eigenvalues and thermodynamic quantities,
as functions of $\theta$, must be symmetric about
$\theta=0$, hence they are functions of the quantity
$\alpha(\theta)$ defined by
\beqn
\alpha(\theta)={|\theta|\over\pi}\quad\mbox{for}
\quad|\theta|\leq\pi\;,
\qquad \alpha(\theta+2\pi)=\alpha(\theta)\;.
\label{alpha_defined}
\eeqn
Note that $\alpha$ is non-analytic in $\theta$ at the boson and
fermion points, but any even polynomial in $\alpha$ is analytic
at the boson point, and any even polynomial in
$1-\alpha$ is analytic at the fermion point.
An example is the exact second virial coefficient
\cite{ASWZ:secvir,JSD:secvir,AC:YG:SO},
\beqn
\label{exsecvir}
A_2=\lambda^2\left[{1\over 4}-{1\over 2}(1-\alpha)^2\right]
=\lambda^2\left[-{1\over 4}+{g\over 2}\right],
\eeqn
with $g$ given as in eq.~(\ref{g-alpha}). Here
$\lambda=\hbar\,\sqrt{2\pi\beta/m}$ is the thermal wavelength,
depending on the mass $m$ and the inverse temperature $\beta$.
A polynomial in $\alpha$ which is analytic in $\theta$ both at
the boson and the fermion point, must be constant, because it is
a periodic polynomial in $\theta$.

The third virial coefficient is analytic at the boson as well as
the fermion point, because it is ``supersymmetric'',
i.e.\ symmetric under the substitution $\alpha\to 1-\alpha$
\cite{SenI,SenII}. Being analytic everywhere, and periodic in
$\theta$ with period $\pi$, it should be expandable as
a rapidly converging power series in $\sin^2\!\theta$.
In fact, very precise numerical calculations \cite{SMJMKO},
see also \nocite{JLASRKB,JMKO,JLAKRKBAS}
{[\ccite{JLASRKB}--\ccite{JLAKRKBAS}]},
in combination with first and second order perturbative calculations
for the complete cluster and virial expansions
{[\ccite{AC:YG:SO}--\ccite{MS:JJMV:IZ:a3}]}, give that
\beqn
A_3=\lambda^4\left[{1\over 36}+{\sin^2\!\theta\over 12\pi^2}
-(1.652\pm 0.012)\,10^{-5}\sin^4\!\theta+\cdots\right].
\eeqn

It may be conjectured that all virial
coefficients, with the exception of $A_2$, are analytic functions
of $\theta$. If $A_4$ is analytic, then it must have the form
\beqn
\label{eq:FourierA4}
A_4=\lambda^6\left[{\sin^2\!\theta\over 16\pi^2}
\left({1\over\sqrt{3}}\,
\ln\!\left(\sqrt{3}+2\right)+\cos\theta\right)
+\sin^4\!\theta\left(c_4+d_4\cos\theta\right)+\cdots\right],
\eeqn
where the coefficients of the lowest order terms are fixed
by perturbation theory at the boson and fermion points.
The Monte Carlo results presented here can be fitted to this form
with two parameters,
\beqn
c_4 = -0.0053\pm 0.0003\;,\qquad
d_4 = -0.0048\pm 0.0009\;.
\eeqn
We estimated the $\chi^2$ per degree of freedom (DOF) to be 3.35
for the two-parameter least squares fit, and 30 for the no-parameter
fit to the minimal Fourier series with $c_4=d_4=\ldots=0$. The error
estimation is discussed in Section \ref{MCresults} below.
We have assumed that the systematic errors can be neglected.

In addition to our numerical results, we derive the following
polynomial approximations for the cluster coefficients $b_N$,
valid for any $N$,
\beqn
\label{eq:polapprbN}
\lambda^2\wt{b}_N
={(-1)^{N-1}\over N^2}\prod_{k=1}^{N-1}
\left(1-{N(1-\alpha)^2\over k}\right)
={1\over N^2}\prod_{k=1}^{N-1}
\left(1-{Ng\over k}\right).
\eeqn
One nice property of these polynomials is that
they are analytic functions of $\theta$ at the fermion point,
as the exact cluster coefficients must be. However, they do not give
the correct second derivatives at the boson and fermion points,
known from perturbation theory, although they do give the correct
first derivatives. An alternative way to introduce the same
polynomials is to postulate that the second virial coefficient
is given by eq.~(\ref{exsecvir}), while all higher virial
coefficients are independent of
$\alpha$. That is, these are just cluster coefficients for
two-dimensional exclusion statistics \cite{Serguei},
with the statistics parameter
given by (\ref{g-alpha}). The corresponding second order diagrams
were identified in ref.~\cite{AdV:SO:2}.

In Sections \ref{partf} and \ref{clexp} below we will review
some basic formulae. In Section \ref{MCmethod} we describe
the Monte Carlo method, emphasizing the difference with respect
to ref.~\cite{JMKO}, where an external harmonic oscillator
potential was introduced for regularization. In the method used
here, we regularize by introducing a finite area $A$.
In Section \ref{earesults} we give some analytical results,
either exact or approximate. The Monte Carlo results
are presented in Section~\ref{MCresults}, mostly in the form of
figures. We conclude in Section~\ref{Conclusion} with a few comments.

\section{The $N$-anyon partition function}
\label{partf}

We consider free particles in two dimensions, except that we need
to confine them inside a finite area $A$ for purposes of normalization.
We might use a square box, but it is simpler to use periodic boundary
conditions so that there are no reflecting walls. Note that the periodicity
is only used for normalization, and that when we speak about anyons,
in the path integral formalism, the only restriction is that
the starting points of trajectories should be inside the given area.
Otherwise they propagate freely in the plane and not on the torus.

The $N$-particle partition function is denoted by
$Z_N(\beta)$. In particular,
\beqn
Z_1(\beta)=
\left[\sum_{n=-\infty}^{\infty}\exp\!
\left(-\pi n^2\,{\lambda^2\over A}\right)\right]^2
={A\over\lambda^2}
\left[1+2\sum_{n=1}^{\infty}\exp\!
\left(-\pi n^2\,{A\over\lambda^2}\right)\right]^2\;,
\eeqn
where we have used an identity from ref.~\cite{Bellman}.
Below we will use only the leading term in the limit $A\to\infty$,
that is, we take
\beqn
Z_1(\beta) = {A\over\lambda^2}\;.
\label{Z1area}
\eeqn
The correction terms for finite $A$ are exponential in $A$.
This formula implies the following scaling relation, valid for one
free particle in two dimensions,
\beqn
Z_1(L\beta)={Z_1(\beta)\over L}\;.
\label{Z1scaling}
\eeqn

By definition, a partition of $N$ is a sequence of non-negative
integers, ${\cal P}=(\nu_1,\nu_2,\ldots)$,
such that $N=\sum_{L=1}^{\infty}\nu_L L$.
Let ${\cal C}_N$ denote the set of all partitions of $N$, and let
${\cal C}=\bigcup_{N=0}^{\infty}{\cal C}_N$ and
${\cal C}'=\bigcup_{N=1}^{\infty}{\cal C}_N$.
In this notation we have that
\beqn
\sum_{{\cal P}\in{\cal C}}=
\sum_{N=0}^{\infty}\sum_{{\cal P}\in{\cal C}_N}=
\sum_{\nu_1=0}^{\infty}
\sum_{\nu_2=0}^{\infty}\ldots
\sum_{\nu_L=0}^{\infty}\ldots\;.
\eeqn
A partition ${\cal P}$ of $N$ labels a conjugation class in
the symmetric group $S_N$ of permutations of $N$ particles, in
such a way that $\nu_L$ is the number of cycles of length $L$.
Thus ${\cal C}_N$ may be identified with the set of conjugation
classes in $S_N$.

The partition function for $N$ identical particles
can in general be expanded as a sum over partitions of $N$,
\beqn
\label{eq:ZNcyclegen}
Z_N(\beta)=\sum_{{\cal P}\in{\cal C}_N}
F_{\cal P}(\beta)\,B_{\cal P}(\beta)\;.
\eeqn
Here $B_{\cal P}$ is the contribution from the partition ${\cal P}$
to the free particle bosonic partition function,
\beqn
\label{eq:ZNcycle}
B_{\cal P}(\beta)=\prod_{L=1}^{\infty}{1\over\nu_L!}
\left({Z_1(L\beta)\over L}\right)^{\nu_L}\;,
\eeqn
and the effect of any interaction of the particles,
including the anyonic ``statistics interaction'',
is described by a correction coefficient $F_{\cal P}$.

We consider here the case of free and non-interacting anyons,
with an anyon phase angle $\theta$. In the path integral
representation of the partition function, $F_{\cal P}$
can be interpreted as the generating function
for the probability distribution of winding numbers \cite{JMKO,JM},
\beqn
F_{\cal P}=F_{\cal P}(\beta,\theta)
=\sum_{Q=-\infty}^{\infty}P_{\cal P}(\beta,Q)\,
\exp(-\rmi\theta Q)\;.
\eeqn
$P_{\cal P}(\beta,Q)$ is
the probability of the total winding number $Q$,
given the partition ${\cal P}$ and given the distribution of paths
valid for free bosons at the inverse temperature $\beta$.

Further on, we will label a partition by the set of its addends
[so ${\cal P}=(2,1,0,\ldots)$, which is $4=2+1+1$, will be labelled
211]. By eqs.~(\ref{eq:ZNcyclegen})--(\ref{eq:ZNcycle})
we have in particular that
\beqn
\label{eq:ZFrelgen}
Z_2(\beta)& = &
{1\over 2}\,F_{11}\,Z_1(\beta)^2+
{1\over 2}\,F_2\,Z_1(2\beta)\;,
\nonumber\\
Z_3(\beta)& =&
{1\over 6}\,F_{111}\,Z_1(\beta)^3+
{1\over 2}\,F_{21}\,Z_1(2\beta)Z_1(\beta)+
{1\over 3}\,F_3\,Z_1(3\beta)\;,
\\
Z_4(\beta)& =&
{1\over 24}\,F_{1111}\,Z_1(\beta)^4+
{1\over 4}\,F_{211}\,Z_1(2\beta)Z_1(\beta)^2+
{1\over 8}\,F_{22}\,Z_1(2\beta)^2
\nonumber\\
&&{}+{1\over 3}\,F_{31}\,Z_1(3\beta)Z_1(\beta)+
{1\over 4}\,F_4\,Z_1(4\beta)\;.
\nonumber
\eeqn
Using the scaling (\ref{Z1scaling}),
and writing $Z_N$ for $Z_N(\beta)$,
we get
\beqn
\label{eq:ZFrel}
Z_2& =&
{1\over 2}\,F_{11}\,{Z_1}^2+
{1\over 4}\,F_2\,Z_1\;,
\nonumber\\
Z_3& =&
{1\over 6}\,F_{111}\,{Z_1}^3+
{1\over 4}\,F_{21}\,{Z_1}^2+
{1\over 9}\,F_3\,Z_1\;,
\\
Z_4& =&
{1\over 24}\,F_{1111}\,{Z_1}^4+
{1\over 8}\,F_{211}\,{Z_1}^3+
{1\over 32}\,F_{22}\,{Z_1}^2+
{1\over 9}\,F_{31}\,{Z_1}^2+
{1\over 16}\,F_4\,Z_1\;.
\nonumber
\eeqn
For each partition, $Z_1$ is raised to a power which is the number
of cycles.

\section{The cluster expansion}
\label{clexp}

The grand canonical partition function is
\beqn
\Xi(\beta,\mu)=1+\sum_{N=1}^{\infty}z^N\,Z_N(\beta)\;,
\eeqn
where $z=\exp(\beta\mu)$ is the fugacity and $\mu$ is
the chemical potential. The pressure $P$ is given by the relation
\beqn
\beta P={\ln\Xi\over A}=\sum_{N=1}^{\infty}b_Nz^N\;,
\eeqn
where $b_N$ is the $N$-th cluster coefficient.
An immediate consequence is that
\beqn
\Xi=\prod_{N=1}^{\infty}\exp(Ab_Nz^N)=
\sum_{{\cal P}\in{\cal C}}\prod_{L=1}^{\infty}
{\left(Ab_Lz^L\right)^{\nu_L}\over\nu_L!}\;,
\eeqn
and hence,
\beqn
Z_N=\sum_{{\cal P}\in{\cal C}_N}\prod_{L=1}^{\infty}
{\left(Ab_L\right)^{\nu_L}\over\nu_L!}\;.
\eeqn

We are more interested in the inverse relation, which
follows from the expansion
\beqn
\label{eq:expansionlnXiI}
\ln\Xi
=\sum_{\nu=1}^{\infty}{(-1)^{\nu-1}\over\nu}\left(
\sum_{L=1}^{\infty}z^L\,Z_L\right)^{\nu}
=\sum_{{\cal P}\in{\cal C}'}(-1)^{\nu-1}(\nu-1)!
\prod_{L=1}^{\infty}{\left(z^L\,Z_L\right)^{\nu_L}\over\nu_L!}\;.
\eeqn
Here $\nu\equiv\nu({\cal P})=\sum_{L=1}^{\infty}\nu_L$
is the total number of cycles in the partition ${\cal P}$.
This gives the cluster coefficients in terms of
the $N$-particle partition functions,
\beqn
Ab_N=\sum_{{\cal P}\in{\cal C}_N}(-1)^{\nu-1}(\nu-1)!
\prod_{L=1}^{\infty}{{Z_L}^{\nu_L}\over\nu_L!}\;.
\eeqn
In particular,
\beqn
Ab_1 & = & Z_1\;,\nonumber\\
Ab_2 & = & Z_2-{{Z_1}^2\over 2}\;,\\
Ab_3 & = & Z_3-Z_2Z_1+{{Z_1}^3\over 3}\;,\nonumber\\
Ab_4 & = & Z_4-Z_3Z_1-{{Z_2}^2\over 2}
+Z_2{Z_1}^2-{{Z_1}^4\over 4}\;.\nonumber
\eeqn
In general we may write
\beqn
Ab_N=Z_N+\cdots
=\sum_{{\cal P}\in{\cal C}_N}F_{\cal P}\,B_{\cal P}+\cdots
=Z_1\sum_{{\cal P}\in{\cal C}_N}
G_{\cal P}\,{B_{\cal P}\over{Z_1}^\nu}\;,
\eeqn
where we have introduced new coefficients
\beqn
G_{\cal P}=(F_{\cal P}+\cdots){Z_1}^{\nu-1}\;.
\eeqn
The ``$\cdots$'' in the last formula represents a sum of terms
that are products of ``$F$'' coefficients.

One main point of introducing the ``$G$'' coefficients is that
they tend to a finite limit in the thermodynamic limit $A\to\infty$,
as we will prove below. $G_{\cal P}$ is the ``connected part'' of
$F_{\cal P}$ for any partition ${\cal P}$. The concept of
connectedness will also be made more precise below.

With eqs.~(\ref{Z1area})--(\ref{Z1scaling}), we get that
\beqn
\label{eq:bNcycle}
\lambda^2 b_N=
\sum_{{\cal P}\in{\cal C}_N}
G_{\cal P}\prod_{L=1}^{\infty}{1\over\nu_L!\,L^{2\nu_L}}\;,
\eeqn
and all quantities occurring in this equation are finite in
the $A\to\infty$ limit. In particular,
\beqn
\lambda^2b_2 & =& {G_{11}\over 2}+{G_2\over 4}\;,\nonumber\\
\lambda^2b_3 & =& {G_{111}\over 6}+
{G_{21}\over 4}+{G_3\over 9}\;,\\
\lambda^2b_4 & =& {G_{1111}\over 24}+{G_{211}\over 8}
+{G_{22}\over 32}+{G_{31}\over 9}+{G_4\over 16}\;.\nonumber
\eeqn
We have that $G_1=F_1=1$, $G_N=F_N$ for $N=2,3,4,\ldots$, and
\beqn
G_{11}  & = & (F_{11}-1)Z_1\;,\nonumber\\
G_{111} & = & (F_{111}-3F_{11}+2){Z_1}^2\;,\nonumber\\
G_{21}  & = & (F_{21}-F_2)Z_1\;,\nonumber\\
G_{1111}& = & (F_{1111}-4F_{111}-3{F_{11}}^2
+12F_{11}-6){Z_1}^3\;,\label{GvsF}\\
G_{211} & = & (F_{211}-2F_{21}-F_2F_{11}+2F_2){Z_1}^2\;,\nonumber\\
G_{22}  & = & (F_{22}-{F_2}^2)Z_1\;,\nonumber\\
G_{31}  & = & (F_{31}-F_3)Z_1\;.\nonumber
\eeqn

For bosons and fermions the probability generating functions
can be factorized as
\beqn
\label{eq:factorFP}
F_{\cal P}=\prod_{L=1}^{\infty}{F_L}^{\nu_L}\;,
\eeqn
where $F_L=1$ for bosons and $F_L=(-1)^{L-1}$ for fermions.
This factorization implies that
\beqn
\Xi=\sum_{{\cal P}\in{\cal C}}\prod_{L=1}^{\infty}{1\over\nu_L!}
\left({z^LF_LZ_1(L\beta)\over L}\right)^{\nu_L}
=\prod_{L=1}^{\infty}\exp\left(
{z^LF_LZ_1(L\beta)\over L}\right),
\eeqn
which gives the cluster coefficient
\beqn
b_N={F_NZ_1(N\beta)\over NA}
={(\pm 1)^{N-1}\over N^2\lambda^2}\;.
\eeqn
Thus, for bosons and fermions $G_N=F_N=(\pm 1)^{N-1}$,
while $G_{\cal P}=0$
for every partition ${\cal P}$ containing two or more cycles.

\section{The Monte Carlo method}
\label{MCmethod}

We may use the Monte Carlo method in order to compute numerically
the coefficient $G_{\cal P}$ for a given partition ${\cal P}$
representing a conjugation class in the symmetric group $S_N$.
For that purpose we represent $G_{\cal P}$ as a path integral
over all paths inducing one given permutation represented by ${\cal P}$,
\beqn
\label{eq:Gpathint}
G_{\cal P}Z_1={\cal N}_{\cal P}\int{\cal D}
(\bfit{r}_1(\tau),\ldots,\bfit{r}_N(\tau))
\exp\!\left(-{S\over\hbar}\right)g_{\cal P}\;.
\eeqn
Here $\bfit{r}_j(\tau)$ is the path of particle $j$, as
a function of the imaginary time $\tau$, and $S$ is the free
particle action in imaginary time,
\beqn
S=\sum_{j=1}^N\int_0^{\hbar\beta}\rmd\tau\,\,{m\over 2}\,
\left|\dd{\bfit{r}_j(\tau)}{\tau}\right|^2\;.
\eeqn
We include the Gaussian factor $\exp(-S/\hbar)$ as part of
the integration measure, so that it is the integrand
$g_{\cal P}$ alone that represents the interaction
of the particles, and we include a normalization factor
${\cal N}_{\cal P}$ so that $G_{\cal P}={Z_1}^{\nu-1}$
if $g_{\cal P}=1$ identically.
Note that ${\cal N}_{\cal P}$ is then finite
(i.e.\ $A$ independent), since the path integral is
proportional to ${Z_1}^{\nu}$ when $g_{\cal P}=1$.
Note also that this path integral representation is actually
very general, and can be applied to any $N$-particle system
with (short range) interactions in any dimension, not just
to the $N$-anyon system considered here.

To see what the integrand $g_{\cal P}$ looks like in our case,
let us take the partition 2+1+1 of 4 as an example.
A closed path in the four-particle configuration space
interchanges the positions of two particles, say particles
1 and 2, and takes the remaining two particles back
to their starting points.
The total winding number $Q$ is the sum of six pairwise
winding numbers,
\beqn
Q=Q_{12}+(Q_{13}+Q_{23})+(Q_{14}+Q_{24})+Q_{34}\;.
\eeqn
Note that $Q_{12}$ is an odd integer and $Q_{34}$ an even integer
(remember that the winding numbers are defined such that
a complete revolution corresponds to the winding number 2),
whereas $Q_{13},Q_{23},Q_{14},Q_{24}$ are in general non-integer,
because particles 1 and 2 do not return to their starting positions.
However, the sums $Q_{(12)3}=Q_{13}+Q_{23}$ and
$Q_{(12)4}=Q_{14}+Q_{24}$ are even integers.
Hence $Q$ is an odd integer.
Let $I$ be any subscript, and introduce the notation
\beqn
e_I=1+f_I=\exp(-\rmi\theta Q_I)\;.
\eeqn
In order to compute the coefficient
$G_{211}Z_1=(F_{211}-2F_{21}-F_2F_{11}+2F_2){Z_1}^3$
we take the integrand to be
\beqn
\label{eq:g211diagrexp}
g_{211}& =&
e_{12}\,e_{(12)3}\,e_{(12)4}\,e_{34}
-e_{12}\,e_{(12)3}-e_{12}\,e_{(12)4}-e_{12}\,e_{34}+2e_{12}\nonumber\\
& =&
e_{12}\left(f_{(12)3}\,f_{(12)4}\,f_{34}+f_{(12)3}\,f_{(12)4}
+f_{(12)3}\,f_{34}+f_{(12)4}\,f_{34}\right)\;.
\eeqn
For example, we compute $F_{211}{Z_1}^3$ by integrating
\beqn
\exp(-\rmi\theta Q)=e_{12}\,e_{(12)3}\,e_{(12)4}\,e_{34}\;,
\eeqn
and we compute $2F_{21}{Z_1}^3$ by integrating
\beqn
\exp(-\rmi\theta(Q_{12}+Q_{(12)3}))
+\exp(-\rmi\theta(Q_{12}+Q_{(12)4}))
=e_{12}\,e_{(12)3}+e_{12}\,e_{(12)4}\;.
\eeqn
The second line of eq.~(\ref{eq:g211diagrexp}) may be
represented diagrammatically as
\beqn
G_{211}Z_1=
\mbox{\begin{picture}(40,20)
\put(14,-5){\circle*{5}}
\put(26,-5){\circle*{5}}
\put(20,-5){\circle{12}}
\put(10,15){\circle*{5}}
\put(30,15){\circle*{5}}
\put(10,15){\line(1,0){20}}
\put(10,15){\line(1,-2){10}}
\put(30,15){\line(-1,-2){10}}
\end{picture}}
+\mbox{\begin{picture}(40,20)
\put(14,-5){\circle*{5}}
\put(26,-5){\circle*{5}}
\put(20,-5){\circle{12}}
\put(10,15){\circle*{5}}
\put(30,15){\circle*{5}}
\put(10,15){\line(1,-2){10}}
\put(30,15){\line(-1,-2){10}}
\end{picture}}
+2\mbox{\begin{picture}(40,20)
\put(14,-5){\circle*{5}}
\put(26,-5){\circle*{5}}
\put(20,-5){\circle{12}}
\put(10,15){\circle*{5}}
\put(30,15){\circle*{5}}
\put(10,15){\line(1,0){20}}
\put(10,15){\line(1,-2){10}}
\end{picture}}\;.
\eeqn
The particles are represented as points (filled circles).
The two-cycle is represented by $e_{12}$ in the integrand and by
a circle connecting two particles in the corresponding diagram.
Each factor $f_I$ in 
the integrand is drawn as a single straight line in the diagram.
Note that we should draw {\em labelled} graphs to represent
the four terms in eq.~(\ref{eq:g211diagrexp}).
But since the value of a graph is independent of the labelling,
it is more natural to draw unlabelled graphs and include instead
integer coefficients counting the number of ways each graph
can be labelled. Hence the factor 2 in front of the last graph.

In a similar way we find the diagrammatic representation
\beqn
G_{1111}Z_1=
\mbox{\begin{picture}(40,20)
\put(10,-5){\circle*{5}}
\put(10,15){\circle*{5}}
\put(30,-5){\circle*{5}}
\put(30,15){\circle*{5}}
\put(10,-5){\line(1,0){20}}
\put(10,-5){\line(1,1){20}}
\put(10,-5){\line(0,1){20}}
\put(10,15){\line(1,0){20}}
\put(10,15){\line(1,-1){20}}
\put(30,-5){\line(0,1){20}}
\end{picture}}
+6\mbox{\begin{picture}(40,20)
\put(10,-5){\circle*{5}}
\put(10,15){\circle*{5}}
\put(30,-5){\circle*{5}}
\put(30,15){\circle*{5}}
\put(10,-5){\line(1,0){20}}
\put(10,-5){\line(1,1){20}}
\put(10,-5){\line(0,1){20}}
\put(10,15){\line(1,0){20}}
\put(30,-5){\line(0,1){20}}
\end{picture}}
+12\mbox{\begin{picture}(40,20)
\put(10,-5){\circle*{5}}
\put(10,15){\circle*{5}}
\put(30,-5){\circle*{5}}
\put(30,15){\circle*{5}}
\put(10,-5){\line(1,0){20}}
\put(10,-5){\line(1,1){20}}
\put(10,-5){\line(0,1){20}}
\put(30,-5){\line(0,1){20}}
\end{picture}}
+3\mbox{\begin{picture}(40,20)
\put(10,-5){\circle*{5}}
\put(10,15){\circle*{5}}
\put(30,-5){\circle*{5}}
\put(30,15){\circle*{5}}
\put(10,-5){\line(1,0){20}}
\put(10,-5){\line(0,1){20}}
\put(10,15){\line(1,0){20}}
\put(30,-5){\line(0,1){20}}
\end{picture}}
+4\mbox{\begin{picture}(40,20)
\put(10,-5){\circle*{5}}
\put(10,15){\circle*{5}}
\put(30,-5){\circle*{5}}
\put(30,15){\circle*{5}}
\put(10,-5){\line(1,0){20}}
\put(10,-5){\line(1,1){20}}
\put(10,-5){\line(0,1){20}}
\end{picture}}
+12\mbox{\begin{picture}(40,20)
\put(10,-5){\circle*{5}}
\put(10,15){\circle*{5}}
\put(30,-5){\circle*{5}}
\put(30,15){\circle*{5}}
\put(10,-5){\line(1,0){20}}
\put(10,-5){\line(0,1){20}}
\put(30,-5){\line(0,1){20}}
\end{picture}}\;.
\eeqn
The coefficient in front of each diagram is again the number of
inequivalent ways of labelling the nodes of the graph.
We may also write
\beqn
G_{22}Z_1=
\mbox{\begin{picture}(40,20)
\put(14,-5){\circle*{5}}
\put(26,-5){\circle*{5}}
\put(20,-5){\circle{12}}
\put(14,15){\circle*{5}}
\put(26,15){\circle*{5}}
\put(20,15){\circle{12}}
\put(20,-5){\line(0,1){20}}
\end{picture}}\;,\qquad
G_{31}Z_1=
\mbox{\begin{picture}(40,20)
\put(14,0.5){\circle*{5}}
\put(26,0.5){\circle*{5}}
\put(20,-10){\circle*{5}}
\put(20,-3){\circle{14}}
\put(20,16){\circle*{5}}
\put(20,-3){\line(0,1){19}}
\end{picture}}\;.
\eeqn

We see that only connected diagrams contribute to
the cluster coefficients. It follows that the latter
are finite in the limit $A\to\infty$.
Indeed, any path gives a non-zero contribution to the path
integral represented by some diagram only if for every line
in the diagram, the corresponding winding number is non-zero.
The probability for this to happen for a connected diagram
goes to zero as $(\lambda^2/A)^{\nu-1}$
when $A\to\infty$, since every $L$-cycle path gives a Gaussian
distribution of points which essentially covers only a finite area,
proportional to $\lambda^2$.
Here $\nu$ is the number of cycles, and $\nu-1$ is the minimum
number of links in a connected graph with $\nu$ nodes.
The factor $A^{-\nu+1}$ cancels exactly the divergence of the factor
${Z_1}^{\nu-1}$ included in the definition of $G_{\cal P}$,
eq.~(\ref{GvsF}).

The general meaning of the relations between the $F$ and $G$
coefficients should now be obvious. $F_{\cal P}$ is a sum of both
connected and disconnected diagrams, whereas $G_{\cal P}$ is
the part of the sum including only the connected diagrams.
For example, the relation
\beqn
F_{211}{Z_1}^3 = G_{211}Z_1 + 2G_{21}G_{1}{Z_1}^2
+ G_2G_{11}{Z_1}^2 + G_2G_1G_1{Z_1}^3\;,
\eeqn
which follows from
(\ref{GvsF}), is represented as
\beqn
F_{211}{Z_1}^3=
\underbrace{\rule[-1.5em]{0em}{1em}
\mbox{\begin{picture}(40,20)
\put(14,-5){\circle*{5}}
\put(26,-5){\circle*{5}}
\put(20,-5){\circle{12}}
\put(10,15){\circle*{5}}
\put(30,15){\circle*{5}}
\put(10,15){\line(1,0){20}}
\put(10,15){\line(1,-2){10}}
\put(30,15){\line(-1,-2){10}}
\end{picture}}
+\mbox{\begin{picture}(40,20)
\put(14,-5){\circle*{5}}
\put(26,-5){\circle*{5}}
\put(20,-5){\circle{12}}
\put(10,15){\circle*{5}}
\put(30,15){\circle*{5}}
\put(10,15){\line(1,-2){10}}
\put(30,15){\line(-1,-2){10}}
\end{picture}}
+2\mbox{\begin{picture}(40,20)
\put(14,-5){\circle*{5}}
\put(26,-5){\circle*{5}}
\put(20,-5){\circle{12}}
\put(10,15){\circle*{5}}
\put(30,15){\circle*{5}}
\put(10,15){\line(1,0){20}}
\put(10,15){\line(1,-2){10}}
\end{picture}}
}_{\textstyle G_{211}Z_1}
+
\underbrace{\rule[-1.5em]{0em}{1em}
2\mbox{\begin{picture}(40,20)
\put(14,-5){\circle*{5}}
\put(26,-5){\circle*{5}}
\put(20,-5){\circle{12}}
\put(10,15){\circle*{5}}
\put(30,15){\circle*{5}}
\put(10,15){\line(1,-2){10}}
\end{picture}}
}_{\textstyle 2G_{21}G_1{Z_1}^2}
+
\underbrace{\rule[-1.5em]{0em}{1em}
\mbox{\begin{picture}(40,20)
\put(14,-5){\circle*{5}}
\put(26,-5){\circle*{5}}
\put(20,-5){\circle{12}}
\put(10,15){\circle*{5}}
\put(30,15){\circle*{5}}
\put(10,15){\line(1,0){20}}
\end{picture}}
}_{\textstyle G_2G_{11}{Z_1}^2}
+
\underbrace{\rule[-1.5em]{0em}{1em}
\mbox{\begin{picture}(40,20)
\put(14,-5){\circle*{5}}
\put(26,-5){\circle*{5}}
\put(20,-5){\circle{12}}
\put(10,15){\circle*{5}}
\put(30,15){\circle*{5}}
\end{picture}}
}_{\textstyle G_2G_1G_1{Z_1}^3}
\;.
\eeqn
It is the last term that dominates in the thermodynamic limit,
but it is $G_{211}$ only that contributes to the cluster
coefficient. Thus, as usual, the grand partition function
is a sum of all diagrams but the thermodynamic potential is
a sum of connected diagrams \cite{AGD}.

The Monte Carlo method consists in generating random paths according to
the Gaussian distribution of paths valid for bosons \cite{JMKO}.
Each four-particle path is closed over
the imaginary time interval $\hbar\beta$, in the sense that
the final configuration is identical to the initial one, but
with the particle positions interchanged by a permutation
belonging to the class ${\cal P}\subset S_N$. Consider the partition
$2+1+1=4$, as in the example above. Then particles 1 and 2 should
interchange positions, while particles 3 and 4 should return to
their starting points. We take, arbitrarily, the starting point
for the path of particle 1 to be at the origin, this is then also
the ending point for particle 2. Equivalently, it is the ending point
for particle 1
over the imaginary time interval $2\hbar\beta$. The starting point for
particle 2, equal to the position of particle 1 after half the
imaginary time interval $2\hbar\beta$, can then be generated
according to a Gaussian distribution around the origin.
The starting and ending point for particle 3 is generated
according to a flat distribution inside a square area $A$ centered
on the origin. Similarly for particle 4.

For each four-particle path generated we count the winding numbers
$Q_{12}$, $Q_{(12)3}$, $Q_{(12)4}$, $Q_{34}$ and increment a histogram
$n(Q)$ in the following way. We compute the total
winding number $Q$ and add 1 to $n(Q)$, this takes care of
the integrand $e_{12}\,e_{(12)3}\,e_{(12)4}\,e_{34}$.
We subtract 1 from $n(Q_{12}+Q_{(12)3})$, in order to take care of
the integrand $-e_{12}\,e_{(12)3}$. Similarly, we subtract 1 from
$n(Q_{12}+Q_{(12)4})$ and from $n(Q_{12}+Q_{34})$, and we add 2 to
$n(Q_{12})$.
Finally, $G_{211}$ is the Fourier transform of the histogram $n(Q)$,
multiplied by the normalization factor ${Z_1}^2/n$, where
$n$ is the total number of four-particle paths generated.
The net contribution to the histogram vanishes if
more than one of the three winding numbers $Q_{(12)3}$, $Q_{(12)4}$
and $Q_{34}$ is zero, and this is what ensures a finite limit
as $A\to\infty$ for the computed $G_{211}$.

\section{Exact and approximate polynomials}
\label{earesults}

The first cluster coefficient, with our definition, is
$b_1=1/\lambda^2$.
The exact result for the second cluster coefficient
of free anyons is
\beqn
\lambda^2b_2={G_{11}\over 2}+{G_2\over 4}
={1\over 2}\,(1-\alpha)^2-{1\over 4}\;.
\eeqn
Since $G_{11}$ is even and $G_2$ is odd under the substitution
$\alpha\to 1-\alpha$, this implies that
\beqn
G_{11}=\alpha(\alpha-1)\;,\qquad G_2=F_2=1-2\alpha\;.
\eeqn
It is further known that the third virial coefficient,
\beqn
A_3=-2\,{b_3\over{{b_1}^3}}+4\,{{b_2}^2\over{b_1}^4}\;,
\eeqn
is even under $\alpha\to 1-\alpha$. The odd part of $-2b_3/{b_1}^3$,
which is $-\lambda^4G_{21}/2$, therefore has to
cancel the odd part of $4{b_2}^2/{b_1}^4$,
which is $\lambda^4G_{11}G_2$. This condition
gives another exact result,
\beqn
G_{21}=2F_2\,G_{11}\;.
\eeqn
One further result \cite{DHO}, which is
exact according to the perturbative calculation
of ref.~\cite{dVun}, is
\beqn
F_L=\prod_{k=1}^{L-1}\left(1-{L\alpha\over k}\right)\;.
\eeqn
Our Monte Carlo results for single cycles of length $L\leq 4$
are consistent with this formula, which is a check of our
Monte Carlo calculation as well as of the perturbative calculation.

Another way to compute $G_2=F_2$, $G_{11}$ and $G_{21}$ is to impose
an external harmonic oscillator potential to make the energy spectrum
discrete, and then take the zero frequency limit. Only energy levels
depending linearly on $\alpha$ contribute to these three quantities.
The computation of $G_L=F_L$ for $L>2$ is much more non-trivial, and
in fact the only known method is perturbation theory, because
also states with non-linear $\alpha$ dependence contribute.

This is about as far as one can get with exact results.
However, in the diagrammatic expansions shown above, one may argue
quite generally that the tree graphs are expected to dominate,
because every additional line in a diagram represents another
factor of the type $f_I=\exp(-\rmi\theta Q_I)-1$ in the integrand,
with $Q_I$ an even integer.
This factor vanishes when $Q_I=0$, which will happen with a certain
probability which is definitely non-zero, and even if it does not
vanish it will often have an absolute value smaller than 1.
Furthermore, one may argue that the path integral represented by
a tree graph should approximately factorize in the same way as
its integrand. These two assumptions, of tree diagram dominance
and factorization, lead in a not entirely trivial
way to the following polynomial
approximation for the general coefficient $G_{\cal P}$,
\beqn
\label{eq49}
G_{\cal P}\simeq\wt{G}_{\cal P}=N^{\nu-2}\,{G_{11}}^{\nu-1}\,
\prod_{L=1}^{\infty}(LF_L)^{\nu_L}\;.
\eeqn
There is a factor $F_L$ for every cycle of length $L$, a factor
$L_1 L_2 G_{11}$ for every single line connecting two different cycles of
lengths $L_1$ and $L_2$ (each $L$-factor counts the number of ways
the line can be connected to the cycle), and there is a sum over all
$\nu^{\nu-2}$ possible ways to connect the cycles into a tree graph.
It is perhaps not
obvious how this leads to eq.~(\ref{eq49}); a simple way to understand
the connection is by looking at low order examples: Consider the case of
3 cycles of lengths $L_1$, $L_2$, and $L_3$. They can be connected to
a tree graph in 3 possible ways. This gives a coefficient
\begin{eqnarray*}
  &&F_{L_1} F_{L_2} F_{L_3}
  \left(L_1 L_2 G_{11}\,L_2 L_3 G_{11}+
  L_2 L_3 G_{11}\,L_3 L_1 G_{11}+L_3 L_1 G_{11}\,L_1 L_2 G_{11}
 \right)\\
  &&= (L_1+L_2+L_3)\, {G_{11}}^2\, \prod_{i=1}^3 L_iF_{L_i}, 
\end{eqnarray*}
which agrees with eq.~(\ref{eq49}) since $L_1+L_2+L_3=N$.  
We should point out that eq.~(\ref{eq49}) was first derived
empirically as an approximate representation of the Monte Carlo
results.

Special cases where these polynomial formulae are exact, as
already mentioned, are the cyclic coefficients $G_L=F_L$,
as well as $G_{11}$ and $G_{21}=2F_2\,G_{11}$.
In the three-particle case there is one approximate polynomial,
\beqn
\wt{G}_{111}=3{G_{11}}^2\;.
\eeqn
The four-particle approximate polynomials are:
\beqn
\wt{G}_{1111}=16{G_{11}}^3\;,\quad
\wt{G}_{211}=8F_2{G_{11}}^2\;,\quad
\wt{G}_{22}=4{F_2}^2G_{11}\;,\quad
\wt{G}_{31}=3F_3G_{11}\;.
\eeqn

The polynomial approximations for the $G$ coefficients
imply the polynomial approximations for the cluster coefficients
given in eq.\ (\ref{eq:polapprbN}), and those
imply in turn that the virial coefficients are independent of
the statistics, except for the second coefficient
(see Appendix).
These polynomials (\ref{eq:polapprbN}) are characteristic
of two-dimensional exclusion
statistics with statistics parameter (\ref{g-alpha}).

\section{Monte Carlo results}
\label{MCresults}

In the Monte Carlo integration one has to choose a suitable value
for the dimensionless parameter $A/\lambda^2$, making
a compromise between systematic and statistical errors.
We want the limit $A/\lambda^2\to\infty$, but if we take
$A/\lambda^2$ too large, we get no statistics. As we have seen,
one randomly generated four-particle path contributes to
the numerical path integral only if the paths of all four
particles wind around each other as one cluster. This necessary
condition implies that the systematic errors for large but finite
$A/\lambda^2$ are exponentially small, since the maximum distance
that one particle wanders away from its starting point, has
a Gaussian distribution. The exponential convergence for
the $G$ coefficients is similar to, although not directly related to,
the convergence for $Z_1$.

Exponential convergence means that one should use one
single value of $A/\lambda^2$, as large as practically useful,
rather than use two or three values and extrapolate.
We used the following values,
\beqn
\mbox{partition 1+1+1+1:\ \ \ \ \ \ }\,
&&{1\over 2\pi\,0.05}=\;3.183\;,\\
\mbox{partition 2+1+1:\ \ \ \ \ \ \ \ \ \ }\,
&&{1\over 2\pi\,0.03}=\;5.305\;,\\
\mbox{partitions 2+2 and 3+1:\ }
&&{1\over 2\pi\,0.01}=15.915\;.
\eeqn
Typically, two to five out of 1000 generated four-particle paths gave
non-zero contributions for these partitions.
We have generated $1.7\,10^6$ 
paths for the partition $1+1+1+1$, $16.4\,10^6$ 
for $2+1+1$, $1.5\,10^6$ 
for $2+2$, $17.6\,10^6$ 
for $3+1$, and finally $397\,000$ paths for the partition $4$.

\begin{figure}
\epsfxsize=15.5cm
\epsffile{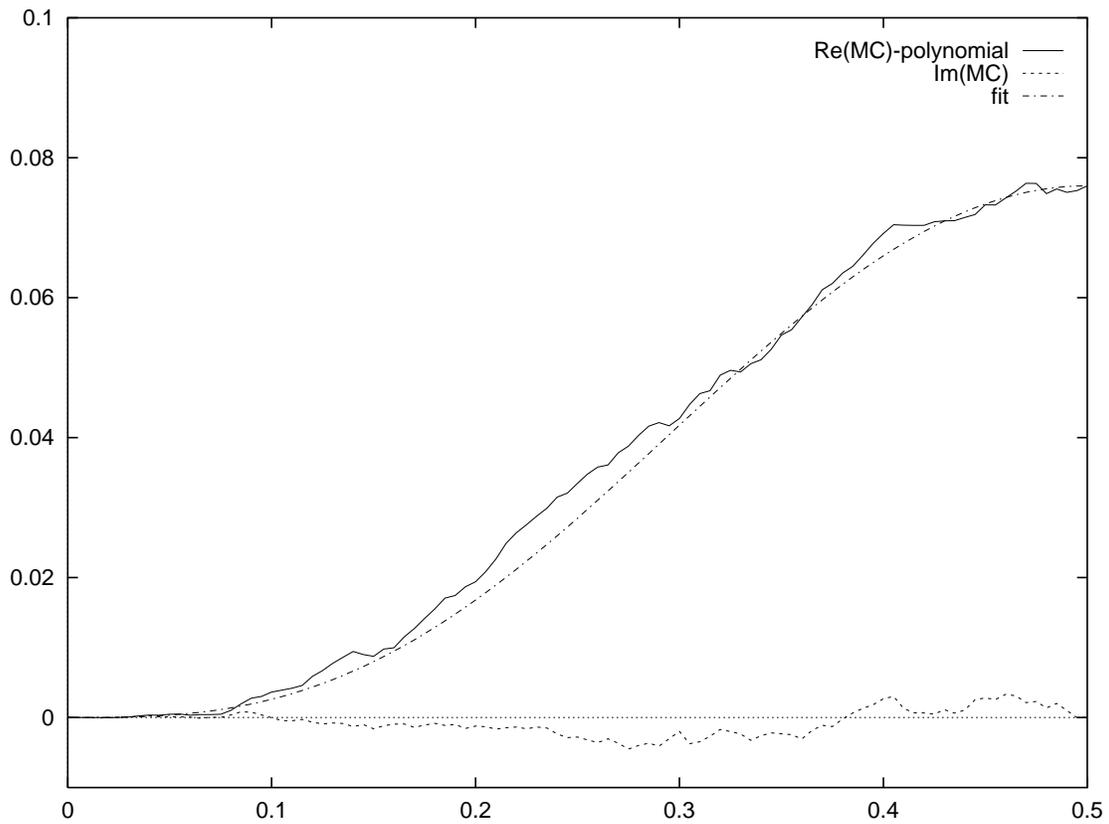}
 
\caption{$G_{1111}-16(\alpha(\alpha-1))^3$
as a function of $\alpha$.
The imaginary part is plotted to indicate the statistical
uncertainty of the real part of the Monte Carlo data.
Only the interval $0\leq\alpha\leq 1/2$ is plotted,
because of the (anti)symmetry about $\alpha=1/2$
The curve marked ``fit'' is
given in eq.~(\ref{eq:fitfig1}).
}
\label{fig:nr1}
\end{figure}

\begin{figure}
\epsfxsize=15.5cm
\epsffile{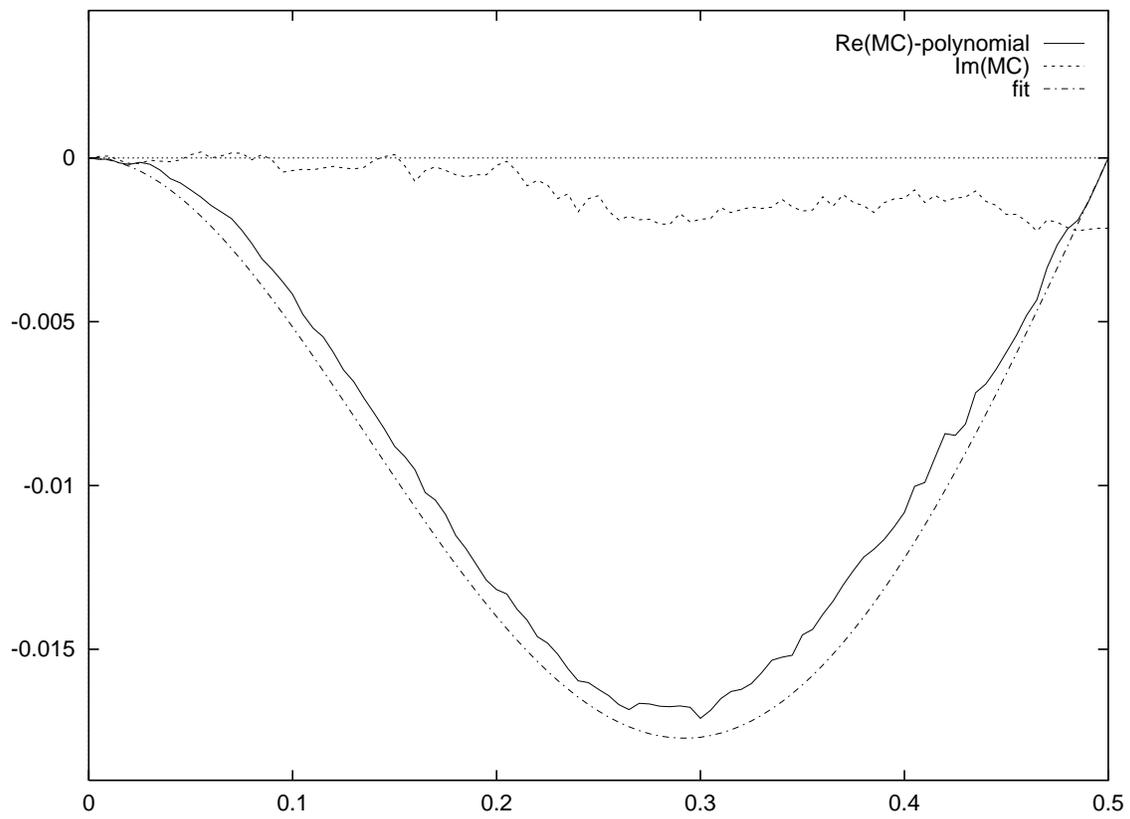}

\caption{$G_{211}-8(1-2\alpha)(\alpha(\alpha-1))^2$
versus $\alpha$.
The curve marked ``fit'' is
given in eq.~(\ref{eq:fitfig2}).
}
\label{fig:nr2}
\end{figure}

\begin{figure}
\epsfxsize=15.5cm
\epsffile{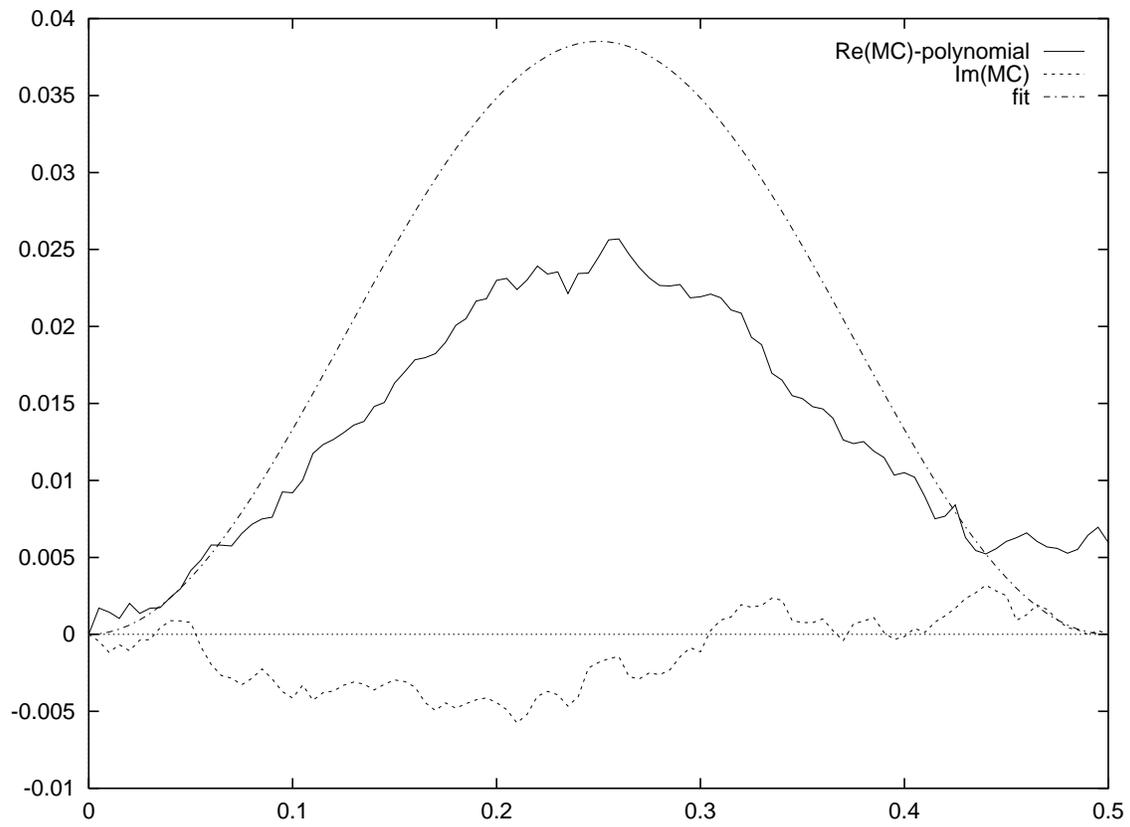}

\caption{$G_{22}-4(1-2\alpha)^2\alpha(\alpha-1)$
versus $\alpha$.
The ``fit'' here is
given in eq.~(\ref{eq:fitfig3}).
}
\label{fig:nr3}
\end{figure}

\begin{figure}
\epsfxsize=15.5cm
\epsffile{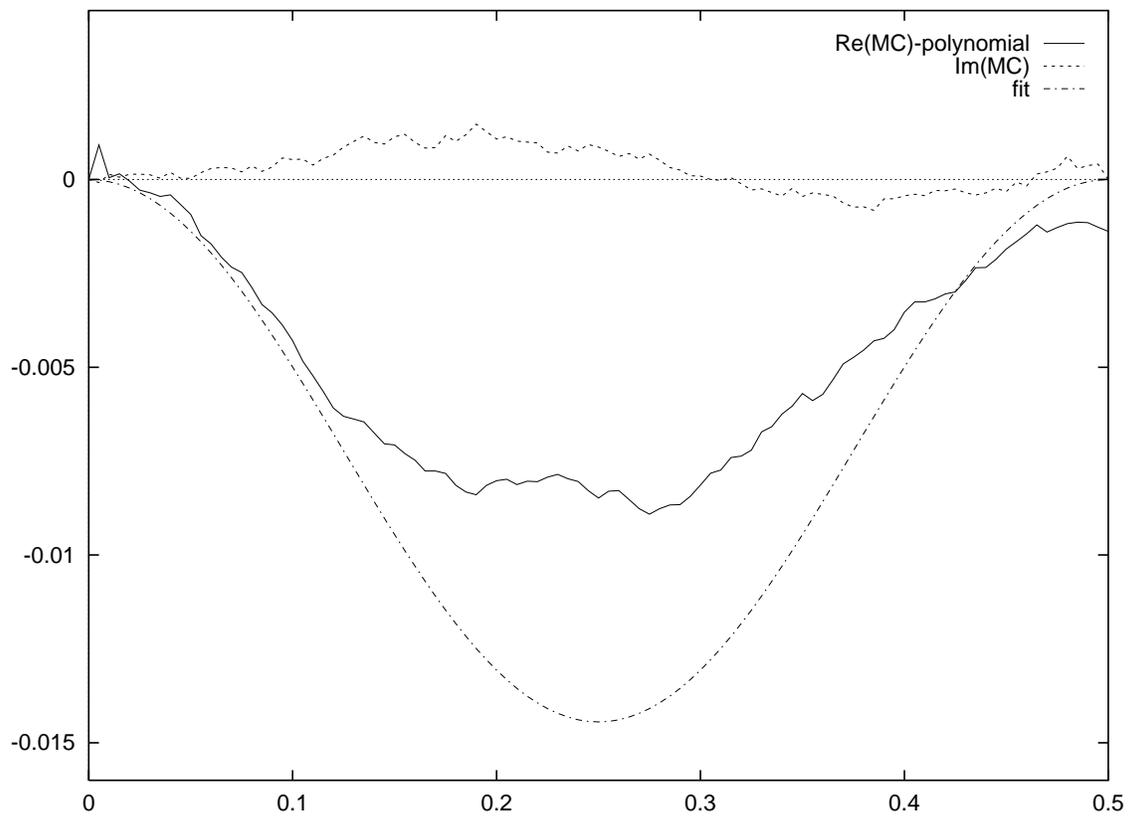}

\caption{$G_{31}-3(1-3\alpha)(1-(3/2)\alpha)\alpha(\alpha-1)$
versus $\alpha$.
The ``fit'' is
given in eq.~(\ref{eq:fitfig4}).
}
\label{fig:nr4}
\end{figure}

\begin{figure}
\epsfxsize=15.5cm
\epsffile{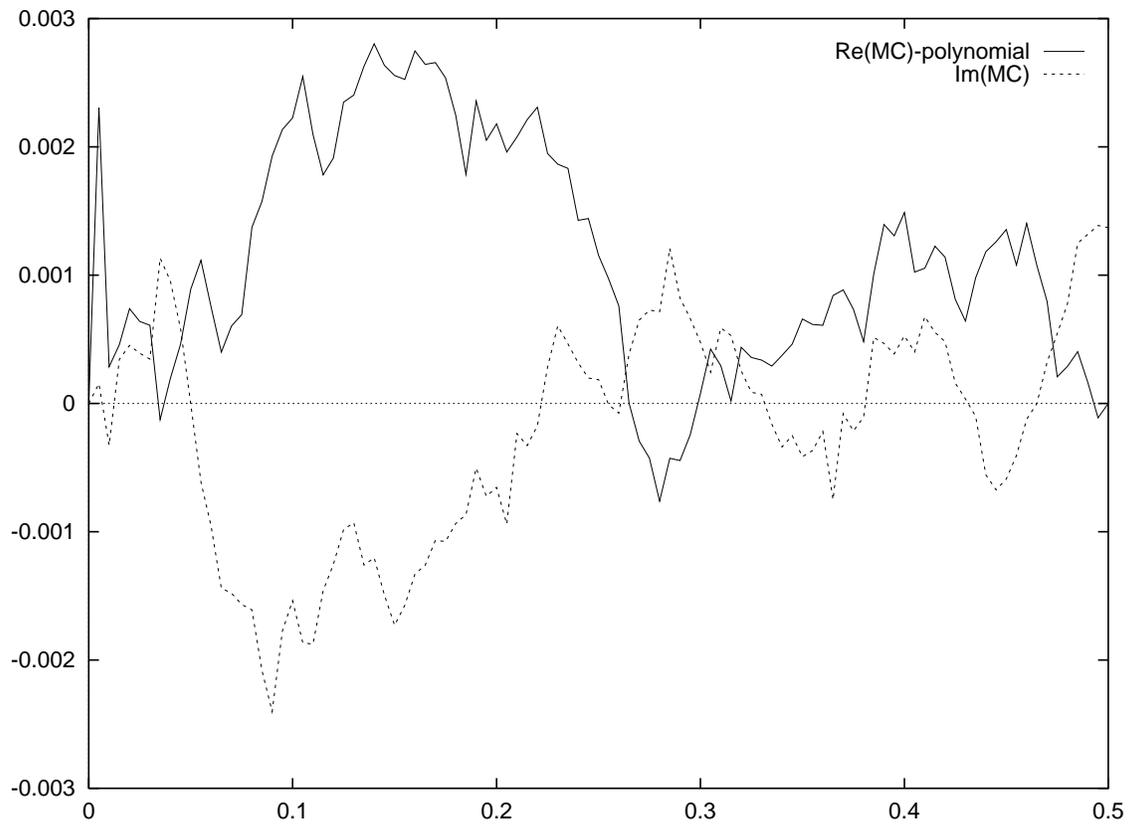}

\caption{$F_4-(1-4\alpha)(1-2\alpha)(1-(4/3)\alpha)$
versus $\alpha$.
}
\label{fig:nr5}
\end{figure}

The computed $G$ coefficients are plotted in the Figures \ref{fig:nr1}
to \ref{fig:nr5}, as functions of $\alpha$. In each case
the polynomial approximation, as discussed in the previous Section,
is subtracted, and the resulting curve is marked
``$\mbox{Re(MC)}-\mbox{polynomial}$''. Because of the statistical
errors, the Monte Carlo generated curve has also a non-zero imaginary
part, marked ``$\mbox{Im(MC)}$'', which is useful because it indicates
the statistical errors in the real part. Since the real part is even
about $\alpha=1/2$ and the imaginary part is odd, or vice versa,
depending on whether the partition is even or odd, we always plot
only the interval $0\leq\alpha\leq 1/2$.

Fig.~\ref{fig:nr1} shows the computed $G_{1111}$ with the polynomial
$16{G_{11}}^3=16(\alpha(\alpha-1))^3$ subtracted.
The curve marked ``fit'' is mostly empirical, and is given by
\beqn
\label{eq:fitfig1}
\mbox{fit}=
-{3\over\pi^2}\,\alpha(\alpha-1)\sin^2\!(\alpha\pi)\;.
\eeqn
The figure shows that this is a perfect fit to the Monte Carlo curve,
within the statistical uncertainty as indicated by the imaginary part.

Fig.~\ref{fig:nr2} shows the computed $G_{211}$ with the polynomial
$8F_2{G_{11}}^2=8(1-2\alpha)(\alpha(\alpha-1))^2$ subtracted.
The curve marked ``fit'' is partly empirical, but with a coefficient
which is chosen so as to produce the correct second order derivative
at $\alpha=0$ \cite{dVun}. The formula is:
\beqn
\label{eq:fitfig2}
\mbox{fit}=
-{2\over 3\pi^2}\;(1-2\alpha)\sin^2\!(\alpha\pi)\;.
\eeqn

Fig.~\ref{fig:nr3} shows the computed $G_{22}$ with
$4{F_2}^2G_{11}=4(1-2\alpha)^2\alpha(\alpha-1)$ subtracted.
The ``fit'' here is
\beqn
\label{eq:fitfig3}
\mbox{fit}=
{2\over\sqrt{3}\pi^2}\,\ln(\sqrt{3}+2)
\sin^2\!(\alpha\pi)\cos^2\!(\alpha\pi)\;.
\eeqn

Fig.~\ref{fig:nr4} shows the computed $G_{31}$ with
$3F_3G_{11}=3(1-3\alpha)(1-(3/2)\alpha)\alpha(\alpha-1)$ subtracted.
Here we have chosen
\beqn
\label{eq:fitfig4}
\mbox{fit}=
{\sqrt{3}\over 4\pi^2}\,\ln(\sqrt{3}+2)
\sin^2\!(\alpha\pi)\cos^2\!(\alpha\pi)\;.
\eeqn

Fig.~\ref{fig:nr5} shows the computed $G_4=F_4$ with the polynomial
$(1-4\alpha)(1-2\alpha)(1-(4/3)\alpha)$ subtracted.
The figure supports the claim that the polynomial is exact.

\begin{figure}
\epsfxsize=15.5cm
\epsffile{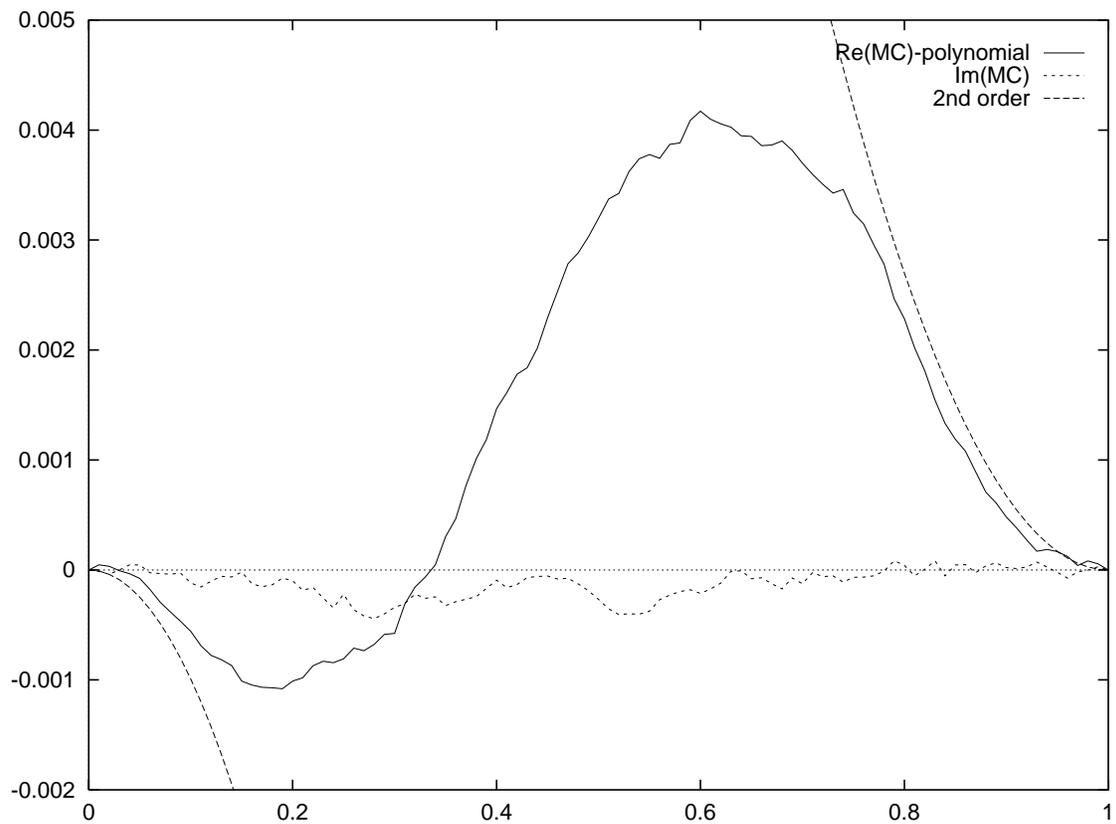}

\caption{The fourth cluster coefficient minus the polynomial
of eq.~(\ref{eq:polapprbN}),
$\lambda^2(b_4-\wt{b}_4)$, as a function of $\alpha$.
Also shown are the parabolas given by the second
order perturbation theory at $\alpha=0$ and $\alpha=1$.
}
\label{fig:nr6}
\end{figure}

Fig.~\ref{fig:nr6} shows the computed cluster coefficient,
$\lambda^2b_4$ with the polynomial $\lambda^2\wt{b}_4$
of eq.~(\ref{eq:polapprbN}) subtracted.
The parabolas given by the second order perturbation theory
at $\alpha=0$ and $\alpha=1$ are shown.

\begin{figure}
\epsfxsize=15.5cm
\epsffile{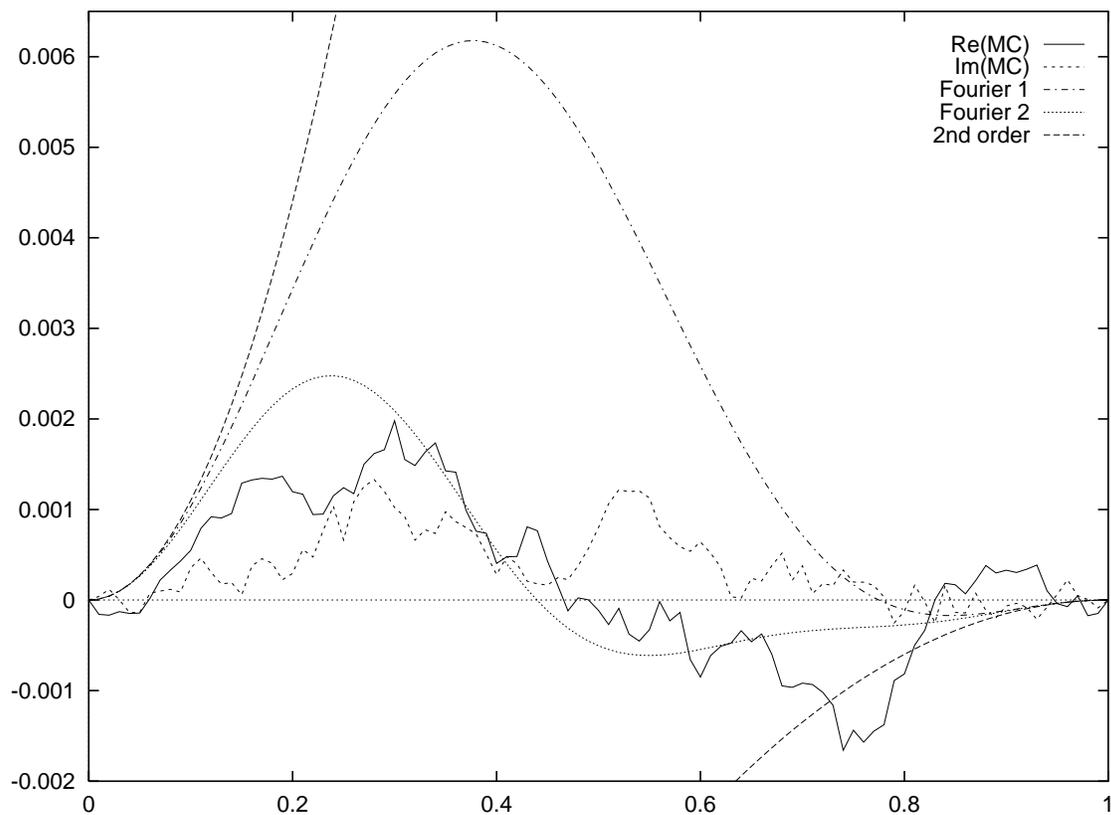}

\caption{The fourth virial coefficient,
$A_4/\lambda^6$, as a function of $\alpha$.
Also plotted are the parabolas given by the second order
perturbation theory at $\alpha=0$ and $\alpha=1$,
and two different Fourier series, as given in
eq.~(\ref{eq:FourierA4}). The curve marked ``Fourier 1''
has $c_4=d_4=\ldots=0$, whereas ``Fourier 2''
is the least squares fit with $c_4=-0.0053$, $d_4=-0.0048$.
}
\label{fig:nr7}
\end{figure}

Fig.~\ref{fig:nr7} shows the computed virial coefficient,
$A_4/\lambda^6$.
The parabolas given by the second order
perturbation theory at $\alpha=0$ and $\alpha=1$ are shown.
Also plotted are two Fourier series, as given in
eq.~(\ref{eq:FourierA4}). The curve marked ``Fourier 1''
has only the two terms required by perturbation
theory, i.e.\ $c_4=d_4=\ldots=0$, whereas the curve marked
``Fourier 2'' is a least squares fit with the
coefficients $c_4=-0.0053$ and $d_4=-0.0048$.

The error estimation for the least squares fit requires some comment.
We neglected the systematic errors, assuming that they are small.
Since the imaginary part of $A_4$ is zero, by definition,
we may estimate the $\chi^2$ per degree of freedom (DOF) for
any fit to the real part of the Monte Carlo data by scaling so that
$\chi^2/\mbox{DOF}=1$ for a comparable fit to the imaginary part.
For the no-parameter fit to $A_4$ with $c_4=d_4=\ldots=0$, we get
$\chi^2/\mbox{DOF}=30$ by comparison with the no-parameter
fit of the exact value zero to the imaginary part.
The same scale factor would give $\chi^2/\mbox{DOF}=1.18$ for
the two-parameter fit to $A_4$ with $c_4$ and $d_4$ as free
parameters. However, the two-parameter fit for the real part should
rather be compared to a two-parameter fit of the same two terms
to the imaginary part, and when we do so, we get instead
$\chi^2/\mbox{DOF}=3.35$.
Comparing the no-parameter and two-parameter fits to the imaginary
part, we find a third scale factor such that these two
degrees of freedom contribute two to $\chi^2$.
The errors of the two fitted parameters $c_4$ and $d_4$ for
the real part are defined as the changes corresponding to
an increase in $\chi^2$ of one, when this third scale factor is
included. This gives $c_4=-0.0053\pm 0.0003$ and
$d_4=-0.0048\pm 0.0009$, with uncorrelated errors.
We arrive at the clear conclusion that the minimal Fourier series,
i.e.\ with $c_4=d_4=\ldots=0$, is excluded by our Monte Carlo
results.

\begin{figure}
\epsfxsize=15.5cm
\epsffile{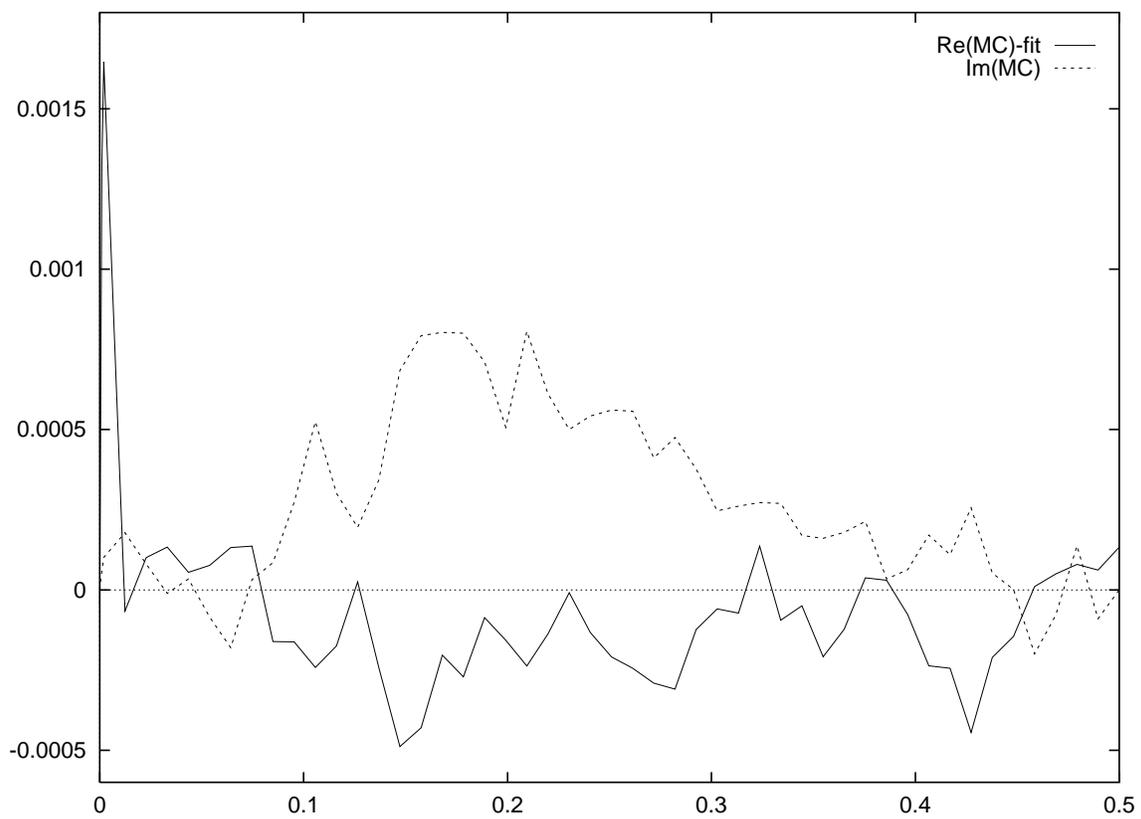}

\caption{$G^{\mbox{\scriptsize line}}_{21}-2\alpha(\alpha-1)-1.07(\alpha(\alpha-1))^2
+0.3(\alpha(\alpha-1))^3$
versus $\alpha$.
}
\label{fig:nr01}
\end{figure}

\begin{figure}
\epsfxsize=15.5cm
\epsffile{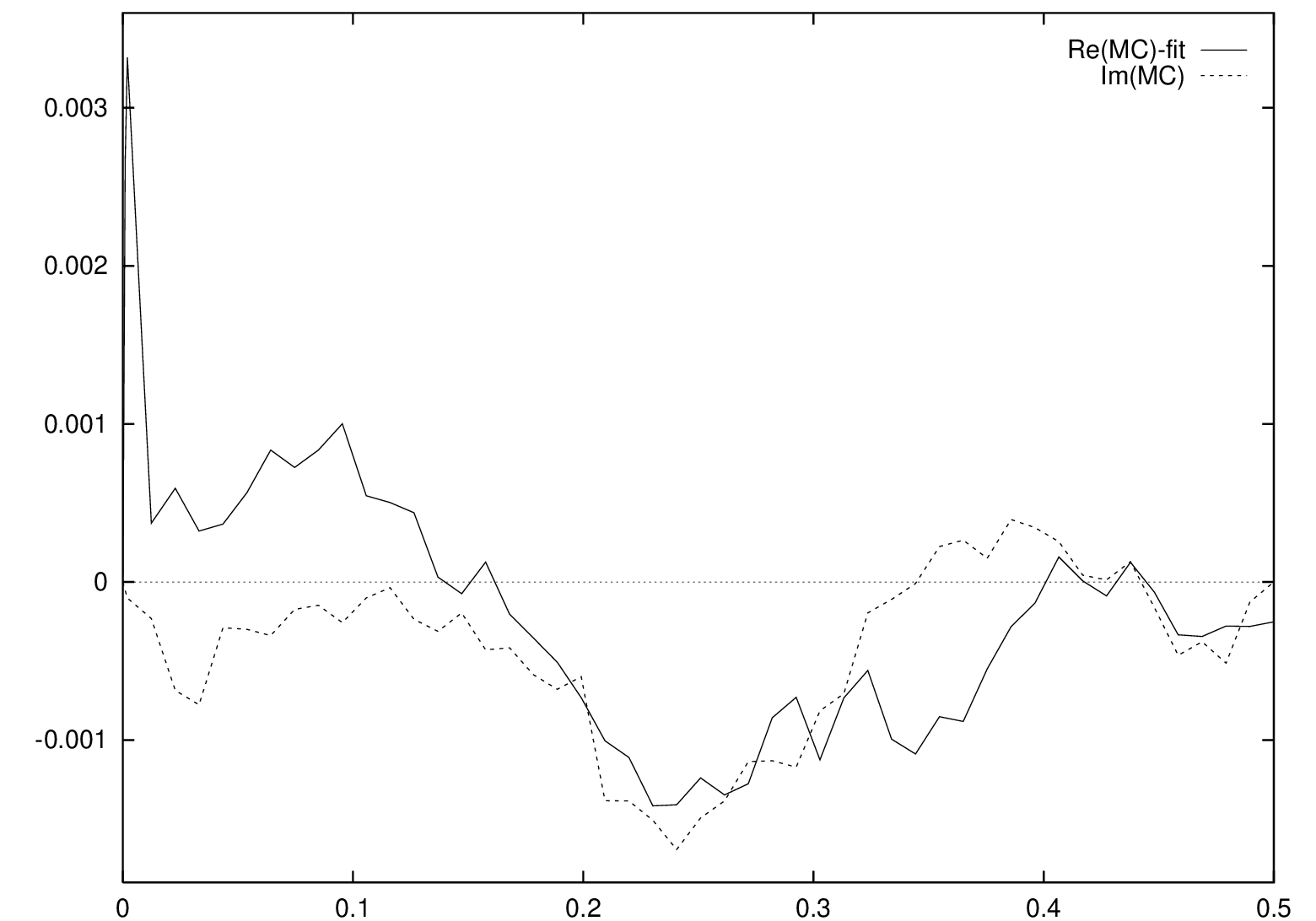}

\caption{$G^{\mbox{\scriptsize line}}_{22}-4\alpha(\alpha-1)-5.77(\alpha(\alpha-1))^2
-2.2(\alpha(\alpha-1))^3$
versus $\alpha$.
}
\label{fig:nr02}
\end{figure}

\begin{figure}
\epsfxsize=15.5cm
\epsffile{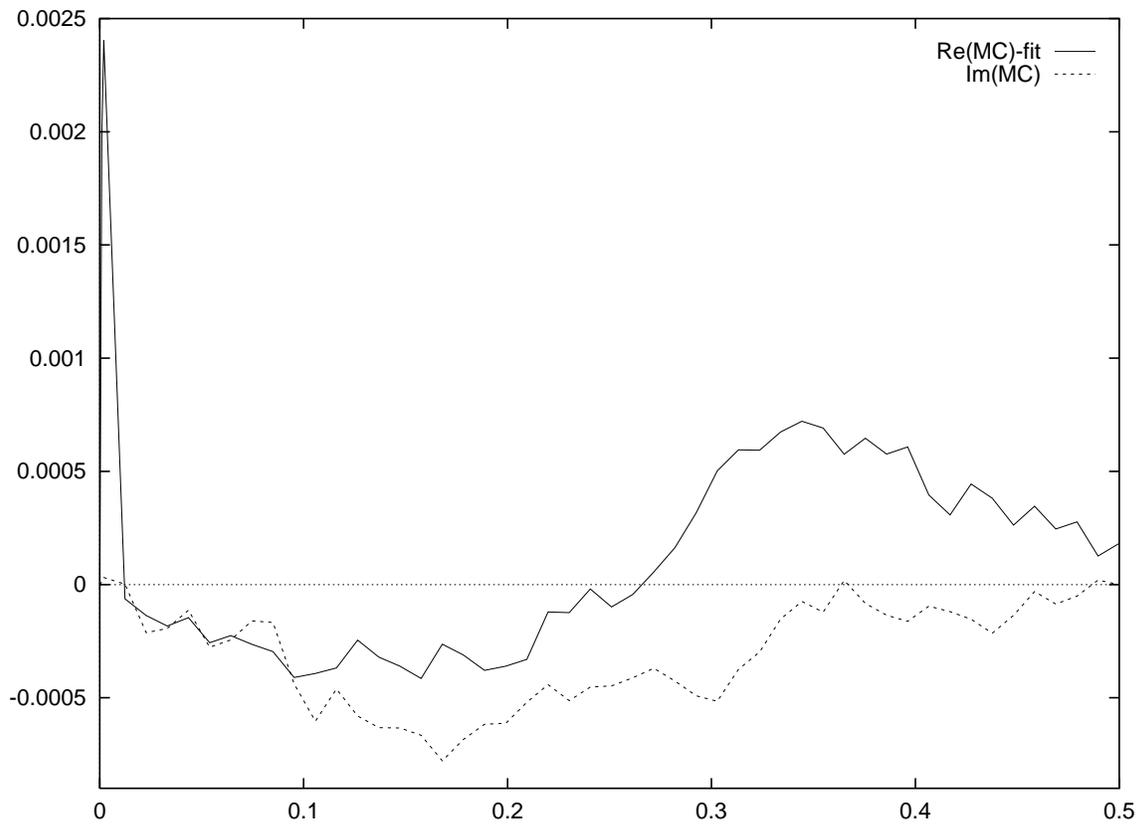}

\caption{$G^{\mbox{\scriptsize line}}_{31}-3\alpha(\alpha-1)-2.46(\alpha(\alpha-1))^2$
versus $\alpha$.
}
\label{fig:nr03}
\end{figure}

We show Figures \ref{fig:nr01}, \ref{fig:nr02} and \ref{fig:nr03}
in order to justify partly the polynomial approximations to
the tree graphs. The quantity plotted is the contribution
$G^{\mbox{\scriptsize line}}_{L_1 L_2}$ to a tree
graph of one single line, joining two cycles of lengths $(L_1,L_2)$,
in the approximation that the integral represented by the graph
factorizes. We claim that this contribution is $L_1L_2G_{11}$,
to lowest order.
Fig.~\ref{fig:nr01} has $L_1=2$ and $L_2=1$, and the ``fit''
which is subtracted from the Monte Carlo data is
\beqn
\mbox{fit}=
2\alpha(\alpha-1)+1.07(\alpha(\alpha-1))^2
-0.3(\alpha(\alpha-1))^3\;.
\eeqn
The important information is the coefficient $L_1L_2=2$ of the
term $G_{11}=\alpha(\alpha-1)$.
Fig.~\ref{fig:nr02} has $L_1=L_2=2$, and the ``fit''
\beqn
\mbox{fit}=
4\alpha(\alpha-1)+5.77(\alpha(\alpha-1))^2
+2.2(\alpha(\alpha-1))^3
\eeqn
has been subtracted.
Fig.~\ref{fig:nr03} has $L_1=3$, $L_2=1$, and we have subtracted
\beqn
\mbox{fit}=
3\alpha(\alpha-1)+2.46(\alpha(\alpha-1))^2\;.
\eeqn
The (unphysical) sharp peaks near $\alpha=0$ in these figures
are due to a deficit
of high winding numbers in the Monte Carlo simulations.

\begin{figure}
\epsfxsize=15.5cm
\epsffile{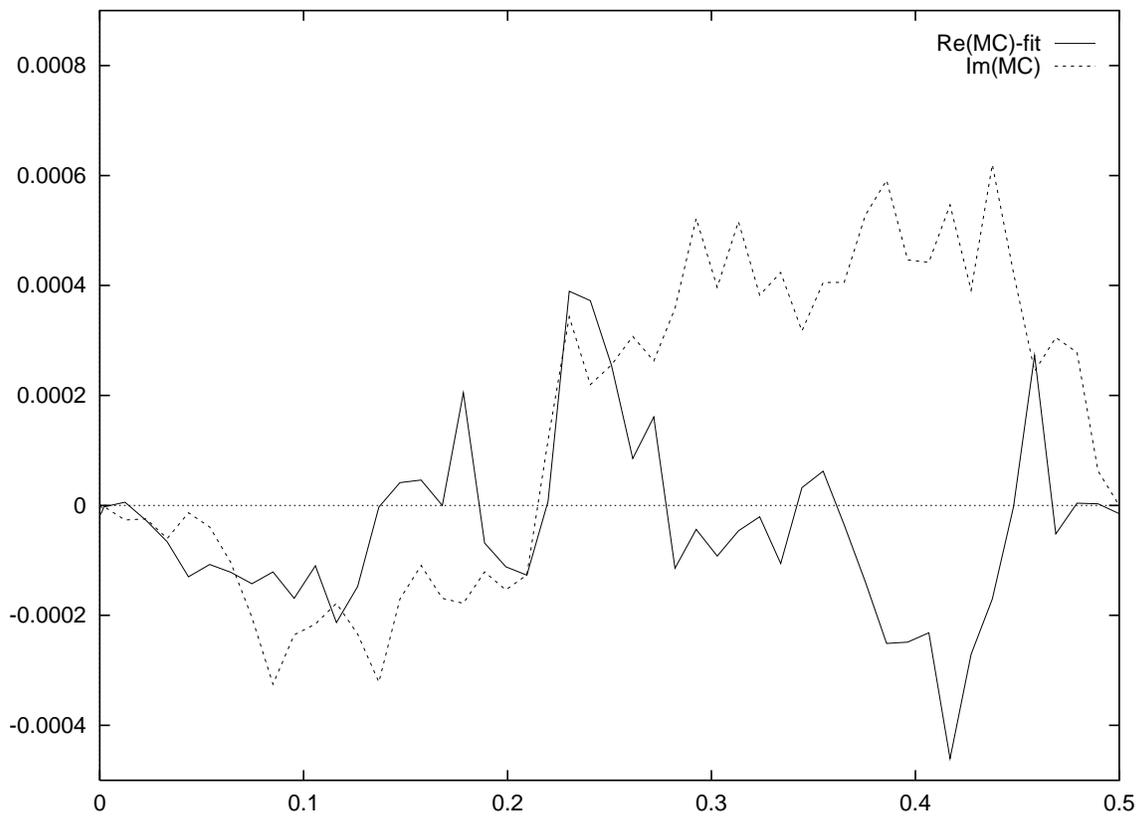}

\caption{The tree graph contribution
$G^{\mbox{\scriptsize tree}}_{111}$
minus the ``fit'' given in eq.~(\ref{eq:fitfig04}).
}
\label{fig:nr04}
\end{figure}

\begin{figure}
\epsfxsize=15.5cm
\epsffile{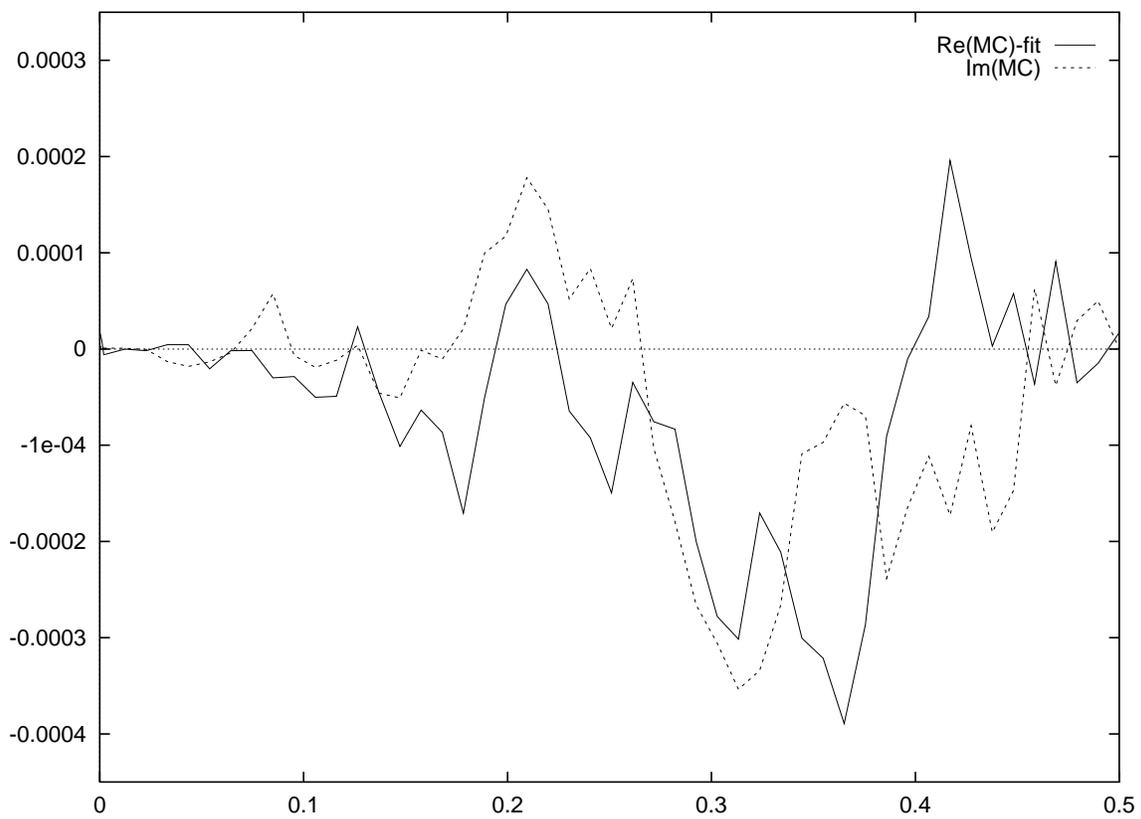}

\caption{The triangle graph contribution
$G^{\mbox{\scriptsize triangle}}_{111}$
minus the ``fit'' given in eq.~(\ref{eq:fitfig05}).
}
\label{fig:nr05}
\end{figure}

The last two figures show the splitting of $G_{111}$ into a tree
graph and a triangle graph, as the simplest possible example of
the contributions of separate graphs. We
Monte Carlo generated $40.7\, 10^6$ three-particle paths at
$A/\lambda^2=1/(2\pi\,0.03)=5.305$.
Fig.~\ref{fig:nr04} shows
the tree graph contribution $G^{\mbox{\scriptsize tree}}_{111}$, after
the following ``fit'' has been subtracted
\beqn
\label{eq:fitfig04}
\mbox{fit}=
3(\alpha(\alpha-1))^2-(1/(4\pi^2))\sin^2\!(\alpha\pi)
-1.41(\alpha(\alpha-1))^3+1.8(\alpha(\alpha-1))^4\;.
\eeqn
Fig.~\ref{fig:nr05} shows
the triangle contribution $G^{\mbox{\scriptsize triangle}}_{111}$, after
subtraction of
\beqn
\label{eq:fitfig05}
\mbox{fit}=
1.41(\alpha(\alpha-1))^3-1.8(\alpha(\alpha-1))^4\;.
\eeqn
This example should give some general idea of what errors
we make by the two approximations of neglecting loop diagrams
and representing the tree graphs by the polynomials of
eq.~(\ref{eq49}).
We find no apparent simple mathematical form for the correction terms.
For the tree graph alone the leading correction to
the polynomial $3(\alpha(\alpha-1))^2$ is of order $\alpha^2$
at the boson point, and $(\alpha-1)^2$ at the fermion point.
The $\sin^2$ form of this
correction, as given in eq.~(\ref{eq:fitfig04}),
reproduces the well known third virial coefficient.
In general, we would expect corrections
of order $\alpha^n$ and $(\alpha-1)^n$ from graphs with $n$ links,
exactly as seen in the present example. However, if we consider not
the perturbative orders in $\alpha$ and $\alpha-1$, but rather
the magnitudes of the maximal corrections from the tree and triangle
graphs, we find that they are comparable.

These considerations suggest that the splitting into graphs
does not simplify the problem of calculating the cluster coefficients.
One would like for example to develop a perturbation theory for calculating
the contributions from separate graphs, but at present we have
no such theory.

\section{Conclusion}
\label{Conclusion}

The main results of this paper are the following.
The fourth virial coefficient of anyons as computed by
the Monte Carlo method is rather well described by
a Fourier series consistent with second-order perturbation theory
and deviating very little from zero.
Our diagram analysis shows in general
that all cluster and virial coefficients are finite, and gives,
in a certain approximation, a direct correspondence with
exclusion statistics. In particular, in this approximation all
the virial coefficients starting from the third are constant.

The above approximation, as well as the fact
that also the exact third and fourth virial coefficients
apparently exhibit quite a regular behavior with $\alpha$,
seem to hint that the interpolation between Bose and Fermi
statistics is rather smooth.
This is in strong contrast
to e.g.\ the anyon superconductivity phenomenon
(non-analytic in $\theta$) believed to
take place at low temperatures {[\ccite{Laughlin1}--\ccite{Chen_et_al}]}.
However, since the dimensionless expansion parameter of the virial series is
$x\equiv\lambda^2\rho$,
which goes to infinity at zero temperature, it is difficult to
predict the low temperature behavior from the first few terms of a
virial expansion.

What we {\em can} test, on the other hand,
is the average field approximation \cite{Wilczek} on which
many of the low temperature predictions are based.
In this approximation one replaces the anyon statistical flux with an
average magnetic field ${\bar B}=\nu\rho\Phi_0$
(with $\Phi_0=h/e$ the elementary quantum of flux),
and treats the anyons as a collection of
non-interacting bosons ($\nu=\theta/\pi$) or fermions ($\nu=\theta/\pi-1$)
in the average magnetic field.
This approximation should work equally well at
high and low temperatures. Thus, we may compare the resulting virial
expansion with our results. Let $f_{\pm}(\rho,B)$ be the
Helmholtz free energy of non-interacting bosons/fermions in
a magnetic field $B$.
Then $f_{\pm}(\rho,\nu\rho\Phi_0)$ is the free energy of anyons
in the average field approximation.
This leads to a virial expansion
\begin{eqnarray}
  \beta P\lambda^2 &=& x \mp \frac{1}{4}\, x^2 +
  \frac{1+3\nu^2}{36}\, x^3 \pm \frac{\nu^2}{16}\,x^4
  -\frac{1-100\nu^2+5\nu^4}{3600}\, x^5 + \cdots
  \label{AverageField}
\end{eqnarray}
Not unexpectedly, eq.~(\ref{AverageField}) does not reproduce
fermions (bosons) correctly when expanded about bosons (fermions).
Implementation of such behavior requires corrections
which have to be put in by hand.
If this is done (by e.g.\ a minimal Fourier series),
it seems that the average field virial
coefficients have a magnitude, and variation
with $\theta$, which up to 4th order in $x$ is qualitatively
rather similar to our results. Since the average field approximation
predicts a thermodynamics with very interesting
low temperature behavior, our results do not rule out that this
will happen in the exact model as well.

\section*{Acknowledgments}

We thank the Centre for Advanced Study, Oslo, where this work was initiated,
for kind hospitality, financial support and use of
computer facilities. This work has received support from the Norwegian
Supercomputer Committee through a grant of computing time.
The work of A.K.\ has been partially funded by the
Norwegian Research Council under contract No.\ 100559/410.
S.M. acknowledges partial support through grants
from the Norwegian Research Council.

\appendix
\section{Polynomials for cluster and virial coefficients}

In this appendix we will prove the result
that the polynomial approximation (\ref{eq:polapprbN}) for
the cluster coefficients
is equivalent to a virial expansion which is the same as for
the two-dimensional free non-relativistic
Bose gas, except that the second virial
coefficient is given by eq.~(\ref{exsecvir}).
We will also prove that the polynomial approximation
(\ref{eq49}) for $G_{\cal P}$ implies eq.~(\ref{eq:polapprbN}).

For simplicity we fix the temperature and choose units such that
$\beta=\lambda=1$.
Thus, e.g., the fugacity is $z=\mbox{e}^\mu$.
We make use of the expansions
\beqn
\rho&=&\dd{P}{\mu}=\sum_{N=1}^{\infty}Nb_N\,z^N\;,
\nonumber\\
\dd{\mu}{\rho}&=&{1\over\rho}\,\dd{P}{\rho}
=\sum_{N=1}^{\infty}NA_N\,\rho^{N-2}\;.
\eeqn
We also define
\begin{equation}
  \rho_g(\mu) = \sum_{N=1}^\infty \frac{z^{N}}{N}\,
  \prod_{k=1}^{N-1}\left(1-\frac{Ng}{k}\right) =
  \sum_{N=1}^\infty z^N\,\frac{(-1)^{N-1}}{Ng}\left(Ng\atop N\right),
  \label{App4}
\end{equation}
which is the density corresponding to the cluster coefficients
$\wt{b}_N$ of eq.~(\ref{eq:polapprbN}). For the Bose gas, with $g=0$,
we have $\rho=-\ln(1-z)$. Shifting the second virial coefficient by
an amount $\Delta A_2=g/2$ then gives
\begin{equation}
  \mu=\ln\left(1-\mbox{e}^{-\rho}\right)+g\rho\;.
  \label{App3}
\end{equation}
For every $g>0$ and every $\mu$, or for $g=0$ and every $\mu<0$,
this equation clearly has a unique solution $\rho>0$.
We want to prove that the solution is $\rho=\rho_g(\mu)$.

For this purpose we rewrite eq.~(\ref{App3}) as
\begin{equation}
  \rho = -\ln\left(1-z \mbox{e}^{-g\rho}\right)= \sum_{n=1}^\infty\,
  \frac{z^{n}}{n}\,\mbox{e}^{-ng\rho}\;,
\end{equation}
and apply the following theorem due to Lagrange \cite{LagrT}:
The equation $\rho = f(\rho)$ has the solution
\begin{equation}
  \rho = \sum_{M=1}^\infty \frac{1}{M!}\,
  {\left.\left(\frac{\rmd}{\rmd r}\right)^{M-1}
  f(r)^{M}\right|}_{r=0}\;.
  \label{App2}
\end{equation}
This gives
\begin{eqnarray}
  \rho &=& \sum_{M=1}^\infty \frac{1}{M!}\,
  \sum_{n_1=1}^\infty \cdots \sum_{n_M=1}^\infty\,
  \frac{z^{n_1+\cdots+n_M}}{n_1\cdots n_M}
  \left.\left(\frac{\rmd}{\rmd r}\right)^{M-1}
  \mbox{e}^{-(n_1+\cdots+n_M)gr}\right|_{r=0}
  \nonumber\\
  &=&\sum_{N=1}^\infty z^{N} \sum_{M=1}^N \left(-Ng\right)^{M-1}\,
  C_{N,M}\;,
  \label{App6}
\end{eqnarray}
where
\begin{equation}
  C_{N,M} = \frac{1}{M!}\sum_{n_1=1}^\infty\cdots
  \sum_{n_M=1}^\infty
  \frac{\delta_{n_1+\cdots+n_M,N}}{n_1\cdots n_M}\,.
  \label{App7}
\end{equation}
What we need to show is that
\begin{equation}
\label{App7b}
   \sum_{M=1}^N (-Ng)^{M-1}\,C_{N,M} =
   \frac{(-1)^{N-1}}{Ng}\left(Ng\atop N\right).
\end{equation}
It is straightforward to show that
\begin{equation}
   \sum_{N=1}^\infty z^N \sum_{M=1}^N g^M C_{N,M} =
   \rme^{-g\ln(1-z)}-1 =
   \sum_{N=1}^\infty (-z)^N \left(-g\atop N\right),
\end{equation}
and hence,
\begin{equation}
\sum_{M=1}^N g^{M-1}\,C_{N,M} =
\frac{(-1)^N}{g}\left(-g\atop N\right).
\end{equation}
Substituting $g\to -Ng$ we get eq.~(\ref{App7b}), completing the proof.

We next turn to the cluster coefficients
\beqn
b'_N=\sum_{{\cal P}\in{\cal C}_N}
N^{\nu-2}\,{G_{11}}^{\nu-1}
\prod_{L=1}^{\infty}{{F_L}^{\nu_L}\over\nu_L!\,L^{\nu_L}}\;,
\eeqn
given by the polynomial approximation in eq.~(\ref{eq49}).
We want to prove that $b'_N=\wt{b}_N$.

We may rewrite the above formula as
\begin{equation}
   b'_N = \sum_{\nu=1}^N
   \frac{N^{\nu-2}\,{G_{11}}^{\nu-1}}{\nu!}
   \sum_{n_1=1}^\infty\cdots\sum_{n_{\nu}=1}^\infty
   \delta_{n_1+\cdots+n_\nu,N}
   \prod_{j=1}^{\nu}\frac{F_{n_j}}{n_j}\;.
   \label{App8}
\end{equation}
To evaluate $\rho=\sum_{N=1}^\infty Nb'_N\,z^N$
we insert eq.~(\ref{App8}),
interchange the summation order of $N$ and $\nu$ and use the relations
$N^{\nu-1}z^N=(\rmd/\rmd\mu)^{\nu-1}z^N$ and
$\sum_{n=1}^{\infty}z^nF_n/n=\rho_{\alpha}(\mu)$. We find
\begin{eqnarray}
  \rho=\sum_{\nu=1}^{\infty} \frac{{G_{11}}^{\nu-1}}{\nu!}\,
  \left(\dd{}{\mu}\right)^{\nu-1}
  \left(\rho_{\alpha}(\mu)\right)^{\nu}
  =\sum_{\nu=1}^{\infty} \frac{1}{\nu!}
  \left.\left(\dd{}{r}\right)^{\nu-1}\,
  \left(\rho_{\alpha}(\mu+G_{11}r)\right)^{\nu}\right|_{r=0}\;.
  \label{App9}
\end{eqnarray}
By the Lagrange theorem, eq.~(\ref{App9}) is the solution to the equation
$\rho = \rho_{\alpha}(\mu+G_{11}\rho)$, which, as
we saw above, is equivalent to
\begin{equation}
   \mu+G_{11}\rho=\ln\left(1-\mbox{e}^{-\rho}\right)+\alpha\rho\;.
\end{equation}
This is precisely eq.~(\ref{App3}) with $g=\alpha-G_{11}=1-(1-\alpha)^2$,
which means that $b'_N=\wt{b}_N$ with $\wt{b}_N$ as given in
eq.~(\ref{eq:polapprbN}).

\newpage

\newpage

\newpage

\newpage

\newpage

\newpage

\newpage

\newpage

\newpage

\newpage

\newpage

\newpage

\end{document}